\documentclass[secnumarabic,amssymb, nobibnotes, aps,prl,twocolumn,longbibliography]{revtex4}

\usepackage{amsmath}
\usepackage{graphicx}
\usepackage{mathtools}
\usepackage{xcolor}
\usepackage{todonotes}
\usepackage[resetlabels,labeled]{multibib}

\newcites{SM}{References for Supplemental Materials}

\begin{document}
\title{Strongly correlated itinerant magnetism on the boundary of superconductivity in a magnetic transition metal dichalcogenide}

\author{Nikola Maksimovic}
\altaffiliation{Contact for correspondence, nikola\_maksimovic@berkeley.edu}

\author{Ryan Day}
\affiliation{Department of Physics, University of California, Berkeley, California 94720, USA}
\affiliation{Materials Science Division, Lawrence Berkeley National Laboratory, Berkeley, California 94720, USA}

\author{Na-Hyun Jo}
\affiliation{Advanced Light Source, Lawrence Berkeley National Laboratory, Berkeley, CA 94720, USA}

\author{Chris Jozwiak}
\affiliation{Advanced Light Source, Lawrence Berkeley National Laboratory, Berkeley, CA 94720, USA}

\author{Aaron Bostwick}
\affiliation{Advanced Light Source, Lawrence Berkeley National Laboratory, Berkeley, CA 94720, USA}

\author{Alex Liebman-Pel\'aez}
\affiliation{Department of Physics, University of California, Berkeley, California 94720, USA}
\affiliation{Materials Science Division, Lawrence Berkeley National Laboratory, Berkeley, California 94720, USA}

\author{Fanghui Wan}
\affiliation{Department of Physics, University of California, Berkeley, California 94720, USA}
\affiliation{Materials Science Division, Lawrence Berkeley National Laboratory, Berkeley, California 94720, USA}


\author{Eli Rotenberg}
\affiliation{Advanced Light Source, Lawrence Berkeley National Laboratory, Berkeley, CA 94720, USA}

\author{Sinead Griffin}
\affiliation{Materials Science Division, Lawrence Berkeley National Laboratory, Berkeley, California 94720, USA}

\author{John Singleton}
\affiliation{National High Magnetic Field Laboratory, Los Alamos National Laboratory, Los Alamos, New Mexico 87545, USA}

\author{James G. Analytis}
\altaffiliation{Contact for correspondence, analytis@berkeley.edu}
\affiliation{Department of Physics, University of California, Berkeley, California 94720, USA}
\affiliation{Materials Science Division, Lawrence Berkeley National Laboratory, Berkeley, California 94720, USA}
\affiliation{CIFAR Quantum Materials, CIFAR, Toronto, Canada}

\begin{abstract}
Metallic ferromagnets with strongly interacting electrons often exhibit remarkable electronic phases such as ferromagnetic superconductivity, complex spin textures, and nontrivial topology. In this report, we discuss the synthesis of a layered magnetic metal NiTa$_4$Se$_8$ (or Ni$_{1/4}$TaSe$_{2}$) with a Curie temperature of 58 Kelvin. Magnetization data and \textit{ab initio} calculations indicate that the nickel atoms host uniaxial ferromagnetic order of about 0.7$\mu_{B}$ per atom, while an even smaller moment is generated in the itinerant tantalum conduction electrons. Strong correlations are evident in flat bands near the Fermi level, a high heat capacity coefficient, and a high Kadowaki-Woods ratio. When the system is diluted of magnetic ions, the samples become superconducting below about 2 Kelvin. Remarkably, electron and hole Fermi surfaces are associated with opposite spin polarization. We discuss the implications of this feature on the superconductivity that emerges near itinerant ferromagnetism in this material, including the possibility of spin-polarized superconductivity. 

\end{abstract}

\maketitle

\section{Introduction}
The discovery of exotic magnetism and superconductivity in Van der Waals heterostructures has brought into focus the urgency of understanding superconductivity on the boundary of itinerant magnetism. The physics of these materials is thought to be driven by strong correlations arising from their flat-band electronic structure, evoking comparisons with heavy fermion materials whose flat-bands arise from the nearly localized $f$-electrons of their magnetic ions~\cite{kang_cascades_2021,kumar_gate-tunable_2022}. The observation of comparable phenomena in both graphene heterostructures and heavy fermion metals, including field re-entrant superconductivity~\cite{cao2021pauli,levy2005magnetic}, possible triplet superconductivity near itinerant magnetism~\cite{Ran2019,zhou2022isospin}, and non-Fermi liquid behavior~\cite{lee2019theory,liu2020tunable} has served to strengthen this connection. Beyond graphene, layered transition-metal dichlagogenides (TMDs) have shown signs of related physics, including the existence of a spin liquid near a Mott insulating phase in 1T'-TaS$_2$~\cite{law_1t-tas2_2017, kumar_gate-tunable_2022}, and time-reversal symmetry breaking superconductivity in 4Hb-TaS$_{2}$~\cite{ribak2020chiral}. Central to the physics of the heavy fermion systems and their exotic superconductivity is the role of correlated itinerant magnetism, but as far as magnetic TMDs are concerned, this physics has not been extensively investigated. The question of whether there are more magnetic TMDs hosting correlated itinerant magnetism near superconductivity, therefore, remains largely open.

In the itinerant picture of ferromagnetism in metals, an imbalance of occupied spin-up and spin-down states in the conduction bands generate a net magnetic moment~\cite{Moriya1984}. Depending on the exchange splitting and the density of majority and minority spin states, the effective magnetic moment can be much smaller than that originating from localized electrons. As a result of this mechanism, perhaps the clearest observable signature of itinerant ferromagnetism is that the ferromagnetic moment in the ordered phase is considerably smaller than that expected of completely localized magnetic dipoles~\cite{takahashi2013spin,rhodes1963effective,moriya1973effect,Moriya1984,Moriya1991}.

On the other hand, there are fundamental outstanding questions in the study of itinerant ferromagnetism especially in cases where the same electrons that host magnetic modes also interact with each other strongly through the Coulomb force, a situation which calls into question effective single-particle band picture described above. This is thought to be the source of much of the exotic physics in correlated itinerant magnets, driving superconducting pairing in the spin-triplet channel mediated by ferromagnetic spin fluctuations~\cite{fay1980coexistence}, a phenomenon which is historically best exemplified in $f$-electron magnets~\cite{hoshino2013itinerant,shen2020strange}. In addition, correlated itinerant ferromagnets often exhibit exotic properties~\cite{fenner2009non,gong2017discovery,huang2017layer,fenner2009non}, including strongly renormalized electron effective masses~\cite{doniach1966low,brinkman1968spin}, topologically non-trivial spin textures~\cite{tonomura2012real}, and non-Fermi liquid behavior~\cite{Brando2016,ritz2013formation,tonomura2012real}, all of which could be related to the emergence of ferromagnetically-mediated superconductivity~\cite{Pogrebna2015,berk1966effect,fay1980coexistence} --- a leading candidate for realizing topological spin-triplet superconductivity in solids~\cite{Ran2019}.

In this study, we present the growth and characterization of a layered metal NiTa$_4$Se$_8$ hosting strong electronic correlations and itinerant ferromagnetism. The presence of strongly interacting electrons is evidenced in calorimetry measurements, resistivity measurements, angle-resolved photoemission, and~\textit{ab initio} calculations. When the nickel concentration in the samples is reduced, the ferromagnetism apparently disappears and the material superconducts at about 2 Kelvin. Moreover, our~\textit{ab initio} calculations suggest that this metal has electron and hole Fermi surfaces, each with opposite spin polarization in the magnetic state --- an unusual manifestation of spin-splitting in a metallic ferromagnet. We discuss possible mechanisms for the superconductivity, and the consequences of electron pairing developing in proximity to itinerant ferromagnetism, potentially in the presence of spin-polarized Fermi surfaces. This study opens the possibility that there may be many more magnetic TMDs whose physics is connected to the heavy fermion unconventional superconductors, but with the advantage of being layered, stable and with potential to be incorporated in highly tunable heterostructures.

\section{Crystal Growth}

Single crystals were grown by a two-step procedure. First, a precursor was prepared. The elements were combined in a ratio Ni:Ta:Se (0.4:1.0:2.0), loaded in an alumina crucible, and sealed in a quartz tube under a partial pressure (200 torr) of Argon gas. The tube was heated to 670$^{o}$C --- the boiling point of selenium --- for 12 hours, and then the temperature was raised to 900$^{o}$C and kept there for 5 days. The furnace was then shut off and allowed to cool naturally. This reaction yields a free-flowing black powder that was ground with a mortar and pestle.

Second, the precursor was loaded with 3 mg/cm$^{3}$ iodine in a 21 cm long quartz tube, evacuated, and placed in a horizontal two-zone furnace. The precursor and iodine were in zone 1 and the other end of the tube (the growth zone) were in zone 2. Both zones were heated to 850$^{o}$C for 3 hours to encourage nucleation. Then, zone 2 was kept at 850$^{o}$C while zone 1 was reduced to 700$^{o}$C. This condition was maintained for 12 hours to clean the growth zone. Finally, the temperature of zone 1 was raised to 850$^{o}$C and that of zone 2 was lowered to 700$^{o}$C. This growth condition was maintained for 5 days after which the furnace was shut off and allowed to cool naturally. Shiny hexagonal crystals up to 5 mm in lateral length were collected from the cold zone. They are easily exfoliated with a scalpel or scotch tape. Crystals of Ta$_{2}$NiSe$_{5}$, which are easily distinguished from NiTa$_{4}$Se$_{8}$ by both color and morphology, were also present in the growth zone.

Energy dispersive X-ray spectroscopy detects an elemental ratio of 0.25:1.00:1.89 (Ni:Ta:Se), suggesting that the samples used in this study are about 5\% selenium deficient. We believe that this deficiency arises due to the vaporization of selenium during the precursor reaction, which could potentially be adjusted for by adding 5-10\% excess selenium to the first stage of the growth procedure.

\section{Results}
\begin{figure}[!htbp]
\centering
\includegraphics[scale=0.7]{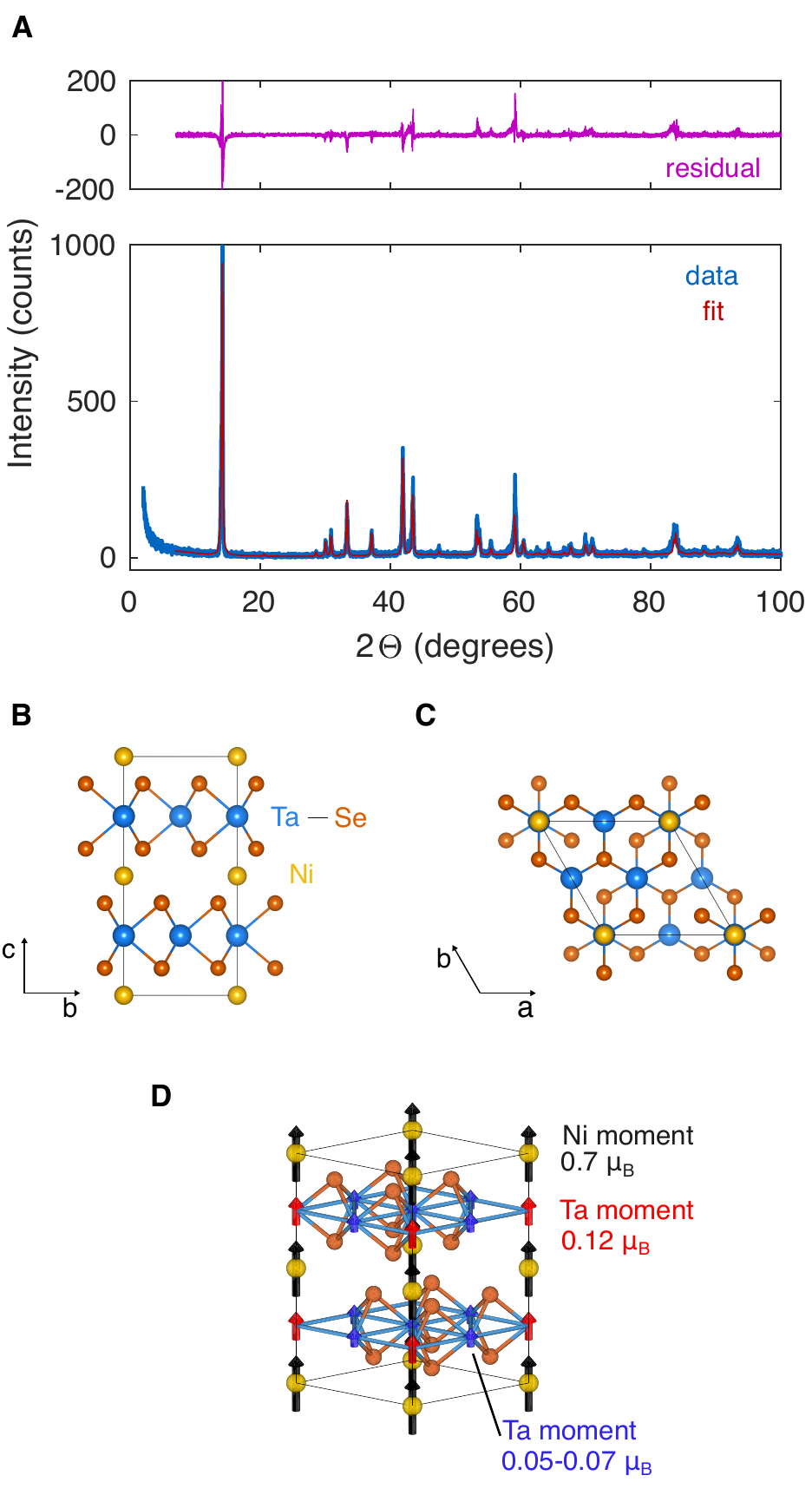}
\caption{{\bf Experimental crystal structure and calculated magnetic structure of NiTa$_{4}$Se$_{8}$} \textbf{A} Powder X-ray diffraction data and refinement (fit shown in red) based on the P63/$mmc$ space group. The residual (the difference between the data and fit) is shown in pink. The lattice parameters extracted from the PXRD refinement are stated in the text. Each prominent peak present in the PXRD data is captured by the refinement based on the crystal structure represented in panels A and B.~\textbf{B} Crystal structure as viewed along the crystallographic $a$-axis. The unit cell is outlined by black lines, and the crystallographic coordinates are shown in the bottom left. Nickel atoms sit between the TaSe$_{2}$ layers.~\textbf{C} Crystal structure viewed along the crystallographic $c$-axis. The tantalum atoms form a triagonal lattice. The nickel atoms similarly form a triangular lattice with twice the periodicity of that of the tantalum atoms.~\textbf{D} Visualization of the magnetic structure resulting from magnetic relaxation density functional theory calculations. The spins are given their own color and labeled by the size of the effective moment per atom.}
\label{fig:pxrd}
\end{figure}

The results of powder X-ray diffraction (PXRD) experiments suggest that NiTa$_{4}$Se$_{8}$ (Ni$_{1/4}$TaSe$_{2}$) crystallizes in the P63/$mmc$ structure. These experiments are performed on precursor powder resulting from a solid-state reaction from which single crystals were grown using chemical vapor transport as described in the crystal growth section. The residual between the fitted PXRD pattern and the experimental one seems to mainly originate from an underestimation in the intensities of the peaks in the refinement as compared to the experiment (Fig.~\ref{fig:pxrd}A). Thus, every significant peak of the experimental PXRD pattern can be accounted for using the crystal structures schematically shown in Figs.~\ref{fig:pxrd}B and C. Based on these diffraction experiments, we conclude that the material is composed of layered basal planes of TaSe$_{2}$ in the 2H polymorphic form, and there are no significant side phases in the precursor powders. NiTa$_{4}$Se$_{8}$ appears to be isostructural to Fe$_{1/4}$TaSe$_{2}$~\cite{morosan2007sharp} and MnTa$_{4}$S$_{8}$~\cite{van1971magnetic}, other doped transition metal dichalcogenides. Between the sheets of TaSe$_{2}$, the nickel atoms sit between the tantalum atoms in the neighboring layers. The nickel atoms themselves form a trigonal layer with twice the $a$-axis periodicity of the tantalum atoms. The crystal structure parameters determined from powder X-ray diffraction refinement are $a = 6.878(6)$\AA~and $c = 12.506(5)$\AA.

Fig.~\ref{fig:pxrd}D shows the results of magnetic relaxation calculations, the details of which are given in the Methods section. We find that all the magnetic moments point along the crystallographic $c$-axis. A moderate magnetic moment of 0.7$\mu_{B}$ per ion lies on the nickel site, while significantly smaller magnetic moments are generated on the tantalum sites --- 0.12$\mu_{B}$ per atom in the tantalum atoms lying directly underneath or above the nickel atoms, and 0.05-0.07 $\mu_{B}$ per atom in the rest of the tantalum atoms. These calculations will prove useful in interpreting the results of physical and magnetic properties measurements.

\begin{figure}[!htbp]
\centering
\includegraphics[scale=0.65]{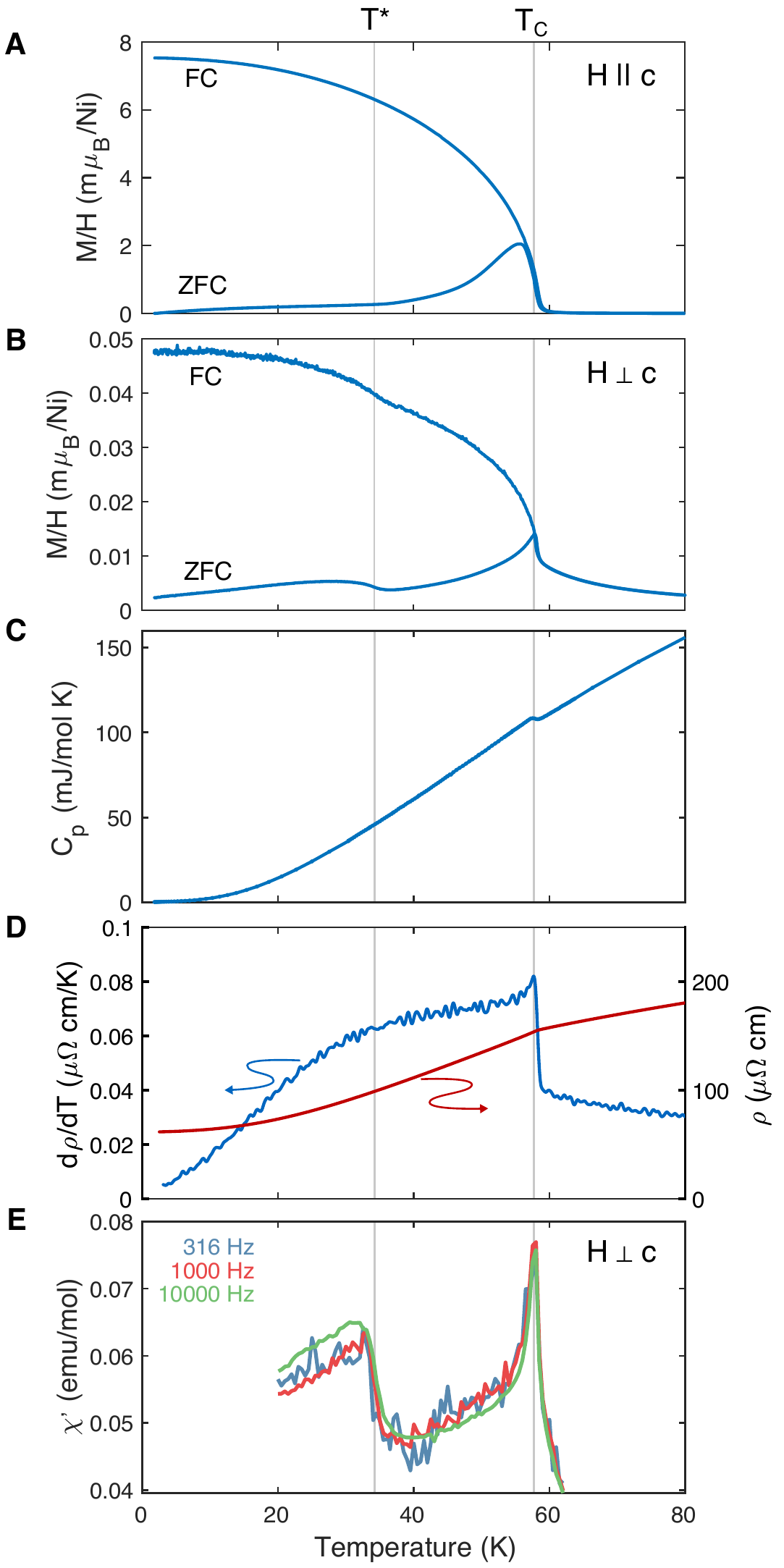}
\caption{{\bf Physical and magnetic properties of NiTa$_{4}$Se$_{8}$} \textbf{A,B} Magnetic susceptibility measured in a field of 100 Oe in the field-cooled (FC) and zero field-cooled (ZFC) protocols with field directed in the interplanar and intraplanar directions.~\textbf{C} Heat capacity.~\textbf{D} Resistance measured in the $ab$ plane and its derivative with respect to temperature. ~\textbf{E} Real ($\chi'$) part of the AC magnetic susceptibility measured with a 10 Oe oscillating field applied in the intraplanar direction.}
\label{fig:phases}
\end{figure}

Fig.~\ref{fig:phases} shows the results of low-temperature physical and magnetic properties characterization experiments on single crystal samples. A spontaneous magnetization develops below a Curie temperature of $T_{C} = 58K$ (Figs.~\ref{fig:phases}A and~\ref{fig:phases}B), as indicated by the splitting of magnetization curves collected with field-cooled and zero field-cooled temperature cycles. In addition, a secondary feature appears in the traces at a temperature of $T^{*} = 36K$, both in the in-plane and out-of-plane directions. This temperature is also associated with a change in curvature in the resistivity-temperature curve of a separate sample (Fig.~\ref{fig:phases}C). $T^{*}$ is also resolved in AC magnetic susceptibility measurements on yet another sample (Fig.~\ref{fig:phases}E) and in magnetization measurements of polycrystalline powder (Fig.~\ref{fig:sup_powdermag}). While the main ferromagnetic-like transition at $T_{C}$ is clearly resolved in heat capacity measurements, $T^{*}$ is not, suggesting that $T^{*}$ might be a phase transition associated with an undetectably small change in entropy, or is not a phase transition at all. The possibility of magnetic impurity phases is addressed in the discussion. Notably, the real part of the AC susceptibility response ($\chi'$) is largely independent of the drive frequency between 300 Hz and 10 kHz, and its overall shape is qualitatively similar to that of the DC magnetic susceptibility response shown in Fig.~\ref{fig:phases}B.

\begin{figure}[!htbp]
\centering
\includegraphics[scale=0.68]{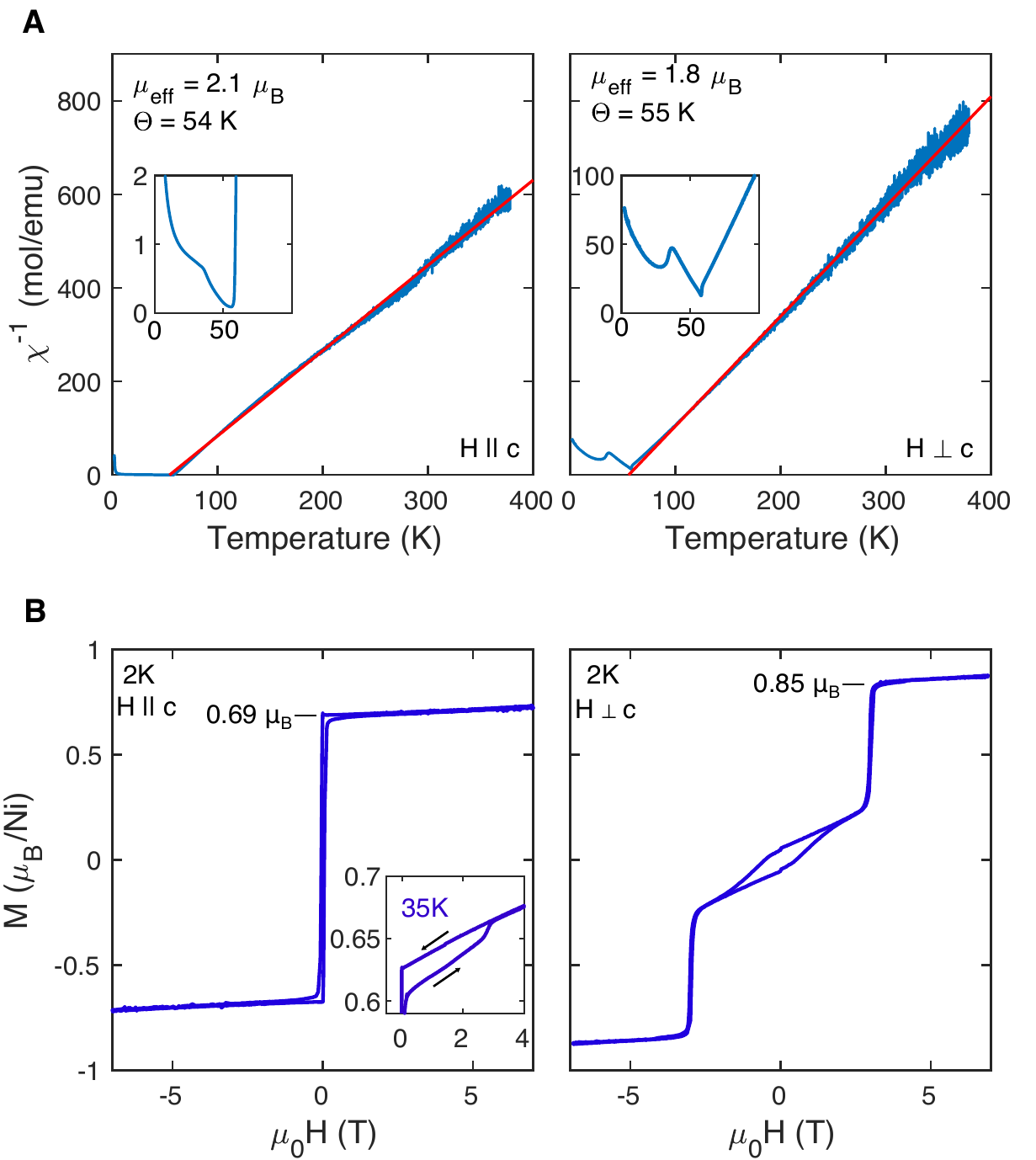}
\caption{{\bf Magnetization measurements and analysis} \textbf{A} Inverse susceptibility versus temperature for the out-of-plane ($H \parallel c$) and in-plane ($H \perp c$) directions. The effective fluctuating moment per nickel atom, $\mu_{\text{eff}}$, extracted from a Curie-Weiss fit (red line) is shown in the panels, along with the Curie temperature $\Theta$. The insets show a zoom in of the inverse susceptibility at low temperature, which shows a double-dip feature.~\textbf{B} Magnetization versus field hysteresis loops for two crystallographic directions. The out-of-plane loop exhibits a sharp coercive field event, with a saturated magnetic moment of $0.69$ bohr magneton per nickel atom. The inset shows that a smaller coercive field event is present at a higher field (see also Fig.~\ref{fig:sup_coercive}). For in-plane magnetic field, there is an apparent metamagnetic transition at a field of about 3 Tesla, and the saturated moment is about 0.85 bohr magneton per nickel.}
\label{fig:magnetization}
\end{figure}

Fig. 3 shows the results of magnetic characterization measurements and analysis. Fig.~\ref{fig:magnetization}A shows the inverse susceptibility with a fit to the Curie-Weiss law ($\chi = C/(T-\Theta)$), where $\Theta$ is the Curie temperature and $C$ is a coefficient proportional to the effective moment ($C = N\frac{\mu_{0}\mu_{\text{eff}}}{3k_{B}}$), where $\mu_{\text{eff}}$ is the effective moment, and $N$ is the concentration of moments in the material. For field in both the in-plane and out-of-plane configurations, a similar Curie temperature is found (54-55K), which agrees well with the observed ordering temperature. With the assumption that the magnetism arises purely from the nickel ions, the fluctuating effective moments are 2.1 $\mu_{B}$/Ni and 1.8 $\mu_{B}$/Ni for the out-of-plane and in-plane configurations, respectively, consistent with the magnetic moment associated with a nickel ion in the Ni$^{2+}$ oxidation state. On the other hand, as shown in Fig.~\ref{fig:magnetization}B, the saturated moment taken from low temperature isothermal magnetization field sweeps is found to be 0.69 and 0.85 $\mu_{B}$ per nickel for the out-of-plane and in-plane directions, respectively. Thus, the saturated moment appears to be considerably smaller than the value of the effective Curie-Weiss moment for both crystallographic directions. Such a disparity is a hallmark of itinerant ferromagnetism~\cite{rhodes1963effective,takahashi2013spin}. Note that for fields in the interplanar direction, a secondary coercive field event associated with a tiny magnetic moment (estimated to be about 0.01~$\mu_{B}$ per formula unit at 35 Kelvin) is observed in the hysteresis loops (Fig.~\ref{fig:magnetization}B inset; see also supplement Fig.~\ref{fig:sup_coercive}). Finally, for magnetic field directed along the intraplanar direction, a sharp jump in the magnetization is observed at 3 Tesla, an observation which is most compatible with the presence of a metamagnetic transition. Electrical conductance measurements up to 60 Tesla for a number of field orientations (Fig.~\ref{fig:sup_PDO}) confirm that there are no further transitions at higher fields, suggesting that this transition separates the ferromagnetic ground state from the in-plane field-polarized state. The temperature-dependence of the coercive field, saturated moment, and metamagnetic transition are explored further in Figs.~\ref{fig:sup_coercive} and~\ref{fig:sup_phasediagrams}.

\begin{figure*}[!htbp]
\centering
\includegraphics[width=\textwidth]{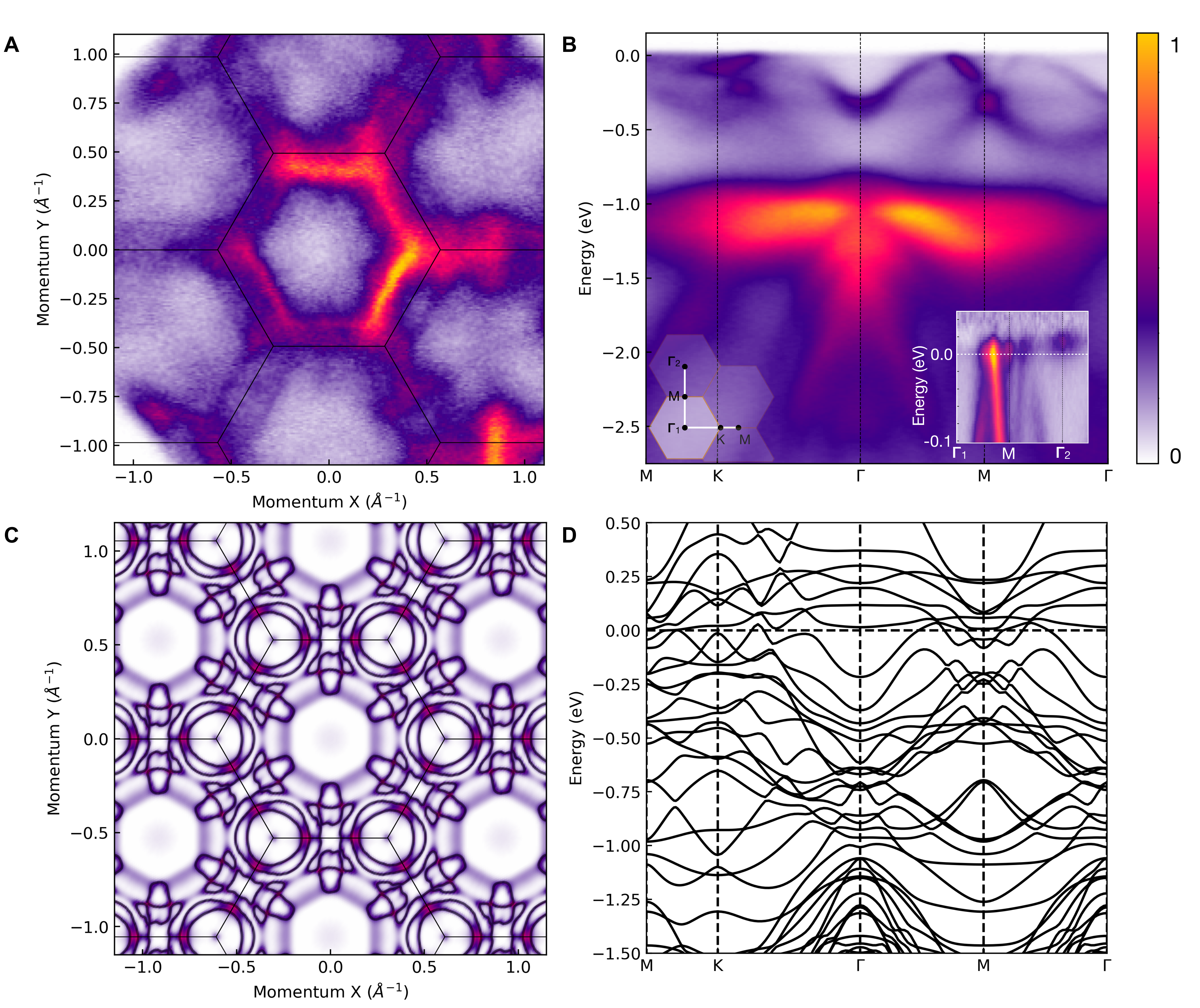}
\caption{{\bf Electronic structure measurements and calculations.} \textbf{A} ARPES-measured Fermi surface, acquired at 11 K. Black contours indicate the boundaries of the Brillouin zone of NiTa$_4$Se$_8$. \textbf{B} ARPES-measured band structure along the high-symmetry path indicated in the inset schematic at 10 K. A high-resolution spectrum near the Fermi level is provided in the inset, displayed after division by the resolution-convolved ($dE=15$ meV) Fermi function. Spectral weight from the thermally-occupied flat bands above $E_F$ is most apparent at $\Gamma_2$. \textbf{C} Two-dimensional Fermi surface projection at $k_z=0$ for NiTa$_4$Se$_8$, as computed by DFT. For closer comparison with ARPES, the Fermi surface (black contours) is overlain along with energy-integrated spectral weight spanning $E_F\pm$20 meV. The hexagonal lattice represents successive neighbouring Brillouin zones. \textbf{D} Band structure along the same high-symmetry path as surveyed by ARPES in \textbf{B}, computed by the same DFT solution as in \textbf{C}.} 
\label{fig:band}
\end{figure*}

Fig.~\ref{fig:band} shows angle-resolved photoemission (ARPES) data in comparison to density functional theory (DFT) calculations of the band structure and fermiology of the NiTa$_{4}$Se$_{8}$. The ARPES and DFT data seem to agree qualitatively well in both the two-dimensional cut of the Fermi surface through the zone center (Figs.~\ref{fig:band}A and C), as well as the band structure (Figs.~\ref{fig:band}B and D). Broadly speaking, the $k_{z}$ cut of the Fermi surface through the zone center consists of two large circular electron-like features centered at the $K$ points, and more complicated hole-like features at the $M$ point. The electron-like pocket at $K$ is reflected in the band structure measurements and calculations shown in Figs.~\ref{fig:band}D and D as a parabolic band crossing the Fermi level, while the complicated set of bands crossing the Fermi level at $M$ produce hole-like pockets. Note also the set of flat electronic bands just above the Fermi level seen in band structure calculations in Fig.~\ref{fig:band}D. The additional spectral  weight near $E=15$ meV in the inset of Fig.~\ref{fig:band}B is likely a signature of these flat bands, indicating their proximity to the Fermi level and therefore potential relevance to the transport and susceptibility functions of NiTa$_{4}$Se$_{8}$.

Consistent with the signatures of flat electron bands in ARPES and DFT calculations, the measured electronic contribution to the heat capacity in NiTa$_{4}$Se$_{8}$ (62 $\pm$ 6 mJ/mol K$^{2}$) is almost as large as some `heavy fermion' $f$-electron metals, which typically have electronic heat capacity coefficients between 100 and 1000 mJ/mol K$^{2}$. From the heat capacity coefficient, we calculate the density of states at the Fermi level to be 13 $\pm$ 1 eV$^{-1}$u.c.$^{-1}$, which is in quite good agreement with that calculated using DFT (11 eV$^{-1}$u.c.$^{-1}$). The $T^{2}$ coefficient of the resistivity, $A$, typically considered to be proportional to the electron-electron scattering rate in Fermi liquids, is 0.032 $\mu\Omega \mathrm{cm/K}^{2}$. These values yield a relatively high Kadowaki-Woods ratio (0.8 $\times 10^{-5}$ $\mu\Omega \mathrm{mol}^{2} \mathrm{K}^{2}/\mathrm{mJ}^{2}$), comparable to that observed in $f$-electron metals~\cite{kadowaki1986universal} and weak itinerant magnets with strong spin fluctuations~\cite{mishra2018spin,brinkman1968spin}. These results suggest that NiTa$_{4}$Se$_{8}$ hosts strong interelectron interactions, even relative to the high density of electronic states at the Fermi level.

\begin{figure*}[!htbp]
\centering
\includegraphics[scale=0.7]{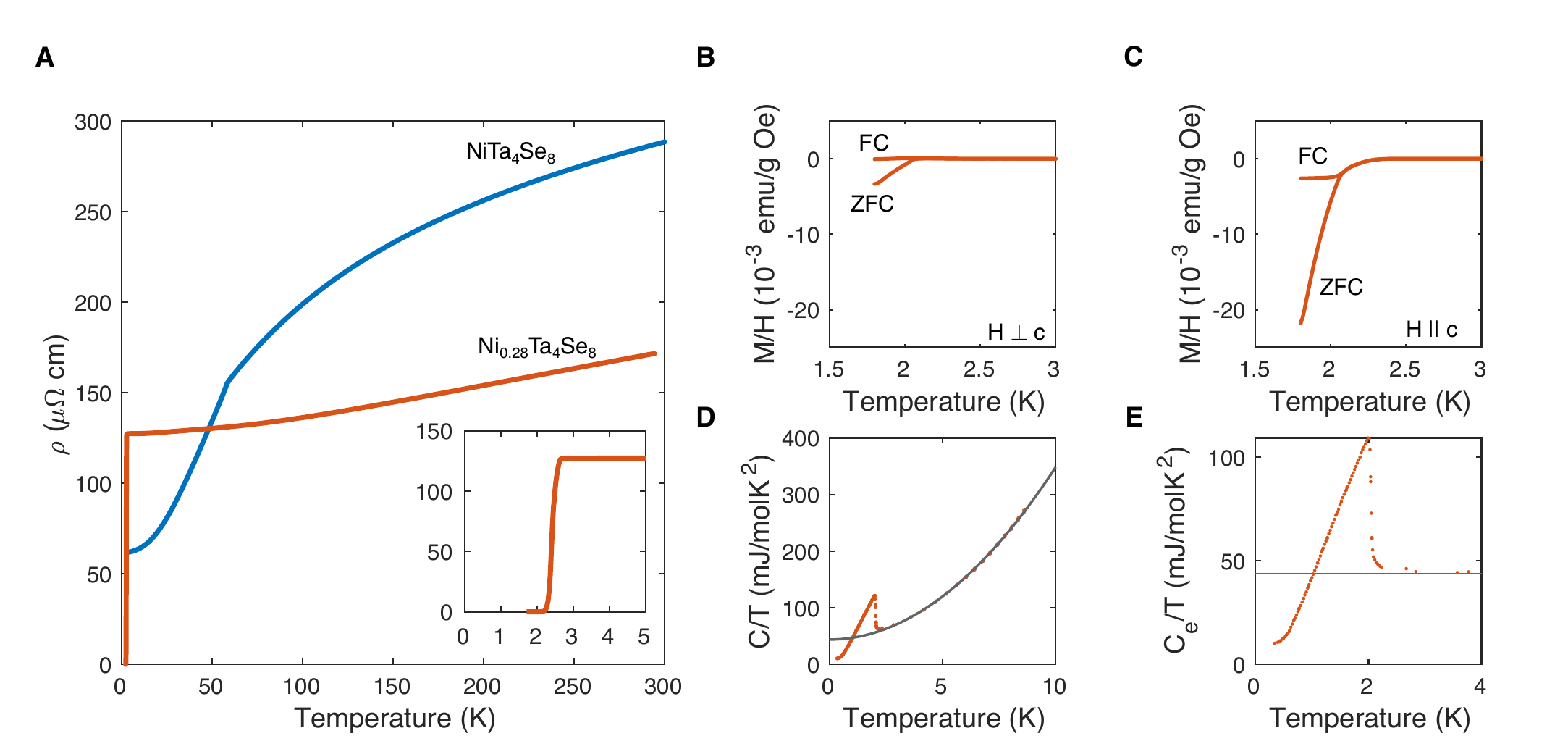}
\caption{{\bf Superconductivity in doped Ni$_{x}$Ta$_{4}$Se$_{8}$} \textbf{A} Resistivity in pure NiTa$_{4}$Se$_{8}$ and doped Ni$_{0.28}$Ta$_{4}$Se$_{8}$ with reduced nickel concentration. A clear anomaly in the trace is observed in the pure sample at the ferromagnetic transition, while the doped sample exhibits a featureless curve until a superconducting transition at low temperature.~\textbf{B,C} Magnetic susceptibility of the doped sample in the field-zooled (FC) and zero field-cooled (ZFC) protocols with a 10 Oe measurement field directed perpendicular and parallel to the crystallographic $c$-axis.~\textbf{D} Total heat capacity divided by temperature. The grey line is a fit to the Debye model $C/T = \gamma + \alpha T^{2}$, where $\gamma$ is the electron contribution and $\alpha$ is the phonon contribution at low temperature.~\textbf{E} Temperature-dependent electronic contribution of the heat capacity, determined by subtracting the phonon term from the data in panel D. A clear supeconducting anomaly is observed. The horizontal grey line indicates the normal state electron contribution to the heat capacity.}
\label{fig:superconductivity}
\end{figure*}

It is often the case that metals with high densities of electronic states and strong interelectron interactions are susceptible to superconducting instabilities~\cite{lee2006doping,pfleiderer2009superconducting}. Fig.~\ref{fig:superconductivity} explores the emergence of superconductivity in samples where then nickel concentration is reduced during the growth procedure~\cite{li2010growth}. Crystallographic measurements on these samples are shown in Fig.~\ref{fig:sup_dopedstructures}, where we determine that the doped samples are comprised of crystalline TaSe$_{2}$ layers with disordered layers of nickel between each layer --- the nickel is randomly distributed in the reduced concentration lattice. In contrast to the pure material where the ferromagnetic transition is associated with a kink in the resistivity trace, the doped samples exhibit a featureless resistivity trace until a superconducting transition at about 2 Kelvin. Magnetic susceptibility and heat capacity measurements show relatively pronounced diamagnetism and a prominent heat capacity anomaly, suggesting that the superconductivity is bulk. Note that a small residual heat capacity appears to persist down to zero temperature even in the superconducting state (Fig.~\ref{fig:superconductivity}E). Measurements of the upper critical field in doped Ni$_{x}$Ta$_{4}$Se$_{8}$, as well as estimations of the superconducting coherence length in comparison to the mean free path, are presented in Fig.~\ref{fig:sup_criticalfields}. The intralayer and interlayer coherence lengths are 6.9$\pm$0.2nm and 13.5$\pm$1.5nm respectively, while the intralayer mean free path is estimated to be roughly 10nm. From these values, and the strongly anisotropic diamagnetic shielding effect seen in Figs.~\ref{fig:superconductivity}B,C, we conclude that the material is a quasi two-dimensional superconductor in the dirty limit.

\begin{figure}[!htbp]
\centering
\includegraphics[width=\columnwidth]{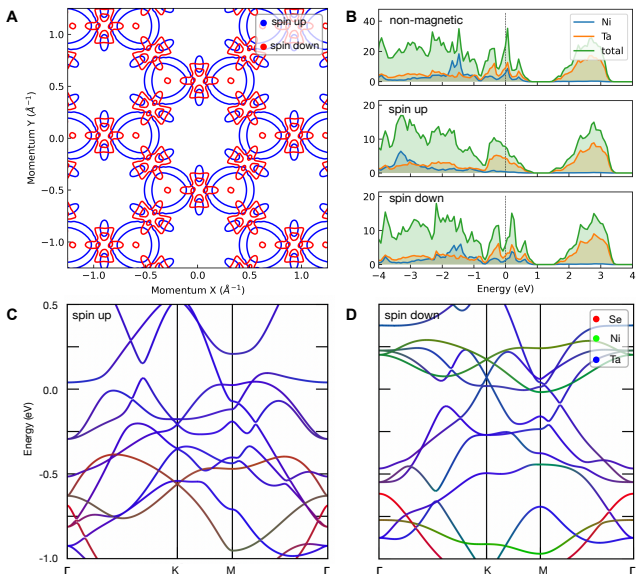}
\caption{{\bf Spin-resolved electronic structure} \textbf{A} Spin-resolved Fermi surface of NiTa$_4$Se$_8$ in the absence of spin-orbit coupling. Spin up and spin down states are indicated in blue and red, respectively. \textbf{B} Density of states for non-magnetic calculation (top), as well as spin-up (middle) and spin-down (bottom) states for spin-polarized calculation. Projection onto Ni and Ta are shown, in addition to the total DOS.~\textbf{C} Bandstructure along high-symmetry path for spin-up states; colours encode atomic projection with Se in red, Ni in green and Ta in blue. \textbf{D} Same as for \textbf{C}, but for spin-down bands.}
\label{fig:spin_dft}
\end{figure}

Spin-resolved calculations of the electronic structure of NiTa$_{4}$Se$_{8}$ in the magnetic state reveal some interesting correlations. First of all, the Fermi surfaces at the $K$ and $M$ points respectively are strongly spin polarized with majority and minority spin electrons, respectively. Given the electronic structure from band-structure calculations and ARPES measurements in Fig.~\ref{fig:band}, we deduce that the electron and hole-like carriers are separately spin up and spin down polarized. There are also qualitative differences in the electronic structures of majority and minority spins in the magnetic state, as summarized in density of states calculations shown in Fig.~\ref{fig:spin_dft}B, and band structure calculations in Figs.~\ref{fig:spin_dft}C and D. The majority spin band centered at the $K$ point has a high dispersion, while the minority spin bands both of Ta $4d$ and Ni $3d$ character are weakly dispersing between the K and M points from above the Fermi level to about 0.25 eV above it --- some shallow dispersive pockets at $K$ and $M$ also exist. This manifests in relatively sharp peaks above and below the Fermi level in the density of electron states specifically of the minority spin species, while the density of states of the majority spins has a smoother profile overall. Such a difference is rare in ferromagnetic materials, where the density of states of two majority and minority spins are expected to be qualitatively similar with an overall energy shift given by the magnetic exchange.

\section{Discussion}

In NiTa$_{4}$Se$_{8}$, we find evidence of itinerant magnetism in that the saturated magnetic moment, at least on the nickel sites, is nearly three times smaller than the effective Curie-Weiss moment. This effect is even more pronounced in the apparently small magnetic moment generated on the tantalum atoms, presumably in the 4$d$ conduction bands with highly itinerant character. Moreover, the material appears to be a strongly correlated metal, as indicated by the peaks in density of states near the Fermi energy (Fig.~\ref{fig:spin_dft}B), the high heat capacity coefficient, and the presence of flat bands above the Fermi energy in band structure calculations and ARPES data (Figs.~\ref{fig:band}D and B). 

The interplay between itinerant magnetism and electron correlations together may be partially responsible for the superconductivity observed in doped samples (Fig.~\ref{fig:superconductivity}) in analogy to superconductivity near correlated magnetism in graphene heterostructures~\cite{cao2021pauli,liu2020tunable} and UTe$_{2}$~\cite{Ran2019}. Other experiments on non-magnetically doped TaSe$_{2}$ and disordered TaSe$_{2}$ suggest that disorder enhances the electron density of states near the Fermi level by suppressing the charge density wave order present in pure 2H-TaSe$_{2}$~\cite{chikina2020turning,li2017superconducting,baek2022interplay}, thereby increasing the superconducting transition temperature. In our case, it is possible that reduction of the nickel concentration brings the sharp peaks in the density of states (Fig.~\ref{fig:spin_dft}B) closer to the Fermi level in a similar vein to previous doping studies of TaSe$_{2}$. However, magnetic impurities are typically expected to suppress, not enhance, superconductivity, making Ni$_{0.28}$Ta$_{4}$Se$_{8}$ fundamentally different from non-magnetically doped variants of TaSe$_{2}$. Given that the nickel is a magnetic dopant and the superconducting state exists in proximity to an itinerant ferromagnetic phase, it is possible that ferromagnetic spin fluctuations are in part responsible for the superconducting pairing~\cite{Ran2019}. The fully spin-polarized nature of the Fermi surfaces at the K points (Fig.~\ref{fig:spin_dft}A) means that electrons in opposite valleys with the same spin could plausibly pair via ferromagnetic fluctuations in a similar manner to that proposed for graphene heterostructures~\cite{liu2020tunable}. 

On the other hand, other non-magnetically doped TaSe$_{2}$ samples are hypothesized to superconduct through an electron-phonon coupling mechanism~\cite{li2017superconducting}. Conventionally the electron-phonon interaction produces even parity coupling which is severely sensitive to magnetic impurities. One possibility is that intervalley pairing supports spin-polarized superconductivity even through an electron-phonon mechanism --- previous theoretical studies have suggested that electron-phonon interactions may support triplet pairing in the presence of strong electronic interactions~\cite{shimahara2004effects} or spin-orbit coupling~\cite{fal2006triplet}. Another related possibility is that the nickel layers induce triplet correlations in the TaSe$_{2}$ electrons through a proximity effect in analogy to artifical superconductor/ferromagnet heterostructures~\cite{buzdin2005proximity}, or that the conduction electrons of TaSe$_{2}$ screen the magnetic impurities in doped Ni$_{x}$TaSe$_{2}$ --- this mechanism would result in the development of subgap states near the magnetic impurities when the sample goes superconducting, which could be detected by local tunneling probes~\cite{liebhaber2019yu}.

Aside from the superconductivity and strong correlations, the magnetism itself in NiTa$_{4}$Se$_{8}$ is notable. Specifically, the apparently two magnetic transitions $T_{C}$ and $T^{*}$ (Fig.~\ref{fig:phases}), the bipartite structure in the out-of-plane hysteresis loops (Fig.~\ref{fig:magnetization}B), and the metamagnetic transition for in-plane fields all warrant further discussion. Because $T^{*}$ appears as a rather broad feature (Fig.~\ref{fig:phases}A), and lacks a significant heat capacity anomaly, it is important to address the possibility of magnetic impurity phases as a source of this anomaly. The most likely candidates are NiSe~\cite{umeyama2012synthesis} and NiSe$_{2}$~\cite{yano2016magnetic} both of which have ferromagnetic ordering temperatures near 20K. However, the secondary feature ($T^{*}$) in susceptibility data on our NiTa$_{4}$Se$_{8}$ crystal occurs closer to 36K. In addition, our PXRD data does not show evidence of peaks associated with either NiSe or NiSe$_{2}$, suggesting that any potential impurity phases constitute an undetectably small fraction of the samples. In addition, both resistivity (Fig.~\ref{fig:phases}B) and Hall effect (Fig.~\ref{fig:sup_Hall}) on single crystal samples exhibit crossover features across 36K, and susceptibility data on powder also exhibits an anomaly at this temperature (Fig.~\ref{fig:sup_powdermag}). For these reasons, we believe that $T^{*}$ is likely to be an intrinsic feature of NiTa$_{4}$Se$_{8}$. Finally, the frequency-independence of $T^{*}$ in AC susceptibility measurements (Fig.~\ref{fig:phases}E) implies that $T^{*}$ does not correlate to spin-glass physics which often arises in frustrated magnets~\cite{lachman2020exchange}. 

The simplest interpretation, motivated in part by our \textit{ab initio} calculations, is that the material has two distinct ferromagnetic moments, and $T^{*}$ corresponds to the freezing temperature of the relatively small tantalum moments while $T_{C}$ corresponds to the freezing temperature of the nickel moments. In itinerant magnets, the jump in heat capacity at the ferromagnetic transition is directly proportional to the size of the ordered moment squared~\cite{clinton1975magnetization} --- thus, the small value of the ordered tantalum moment may explain the apparent absence of $T^{*}$ in heat capacity measurements. The bipartite structure of out-of-plane hysteresis loops is also consistent with the magnetic structure put forward in Fig.~\ref{fig:pxrd}D --- based on a phenomenological free energy model with two order parameters corresponding to the nickel and tantalum moments, respectively, we are able to reproduce the qualitative structure in the out-of-plane hysteresis loops (Fig.~\ref{fig:sup_free_energy_coupled}). These simulations assume a single magnetic domain and uniaxial anisotropy, as expected to first order for hexagonal crystals. Note, however, that the bipartite structure in magnetization loops persists above $T^{*}$ (Fig.~\ref{fig:sup_coercive}). In addition, this free energy model does not reproduce the inner hysteresis loop for fields directed in-plane (Fig.~\ref{fig:magnetization}B).

A central feature in the data, which is not captured immediately by the magnetic structure in Fig.~\ref{fig:pxrd}D, is the presence of a metamagnetic transition for fields directed along the hard axis. We suggest that this feature may originate in the electronic structure of the material. Conventionally in easy axis ferromagnets, one expects a field directed perpendicular to the easy axis to induce a gradual rise in the magnetization until a saturated moment is reached (as exemplified in simulations shown in Fig.~\ref{fig:sup_free_energy_coupled}). However, there are cases where a metamagnetic transition has been observed in ferromagnets with strong anisotropy for fields directed perpendicular to the easy axis, perhaps most notably in compounds like URhGe and UTe$_{2}$ where such a transition coincides with re-entrant superconductivity~\cite{mineev2015reentrant,levy2005magnetic,knebel2019field}. In other cases, like the itinerant ferromagnets LuCo$_{3}$, a metamagnetic transition has been observed under similar field configurations relative to the easy axis, and attributed to a field-induced change in occupancy of $d$-electron states which at a critical field causes a redistribution of the majority and minority spin density of states giving a jump in the net magnetization~\cite{neznakhin2020itinerant}. This essential mechanism for ``itinerant electron metamagnetism'' --- that of field-induced changes to the electronic structures of majority and minority spin species~\cite{wohlfarth1962collective} --- has been considered for several decades to describe metamagnetic transitions in a variety of itinerant ferromagnets~\cite{levitin1998effects,shimizu1982itinerant}. The necessary conditions for the presence of itinerant metamagnetism are that the density of states is relatively large and has positive curvature near the Fermi level~\cite{wohlfarth1962collective}, which causes different free energy terms to compete with each other at finite field. These conditions are certainly fulfilled in NiTa$_{4}$Se$_{8}$ according to our spin resolved density of states calculations and heat capacity measurements. Indeed, the overall field-temperature phase diagram shown in supplement Fig.~\ref{fig:sup_phasediagrams} is reminiscent of that of itinerant ferromagnetic systems~\cite{belitz2005tricritical}, where the transition is driven first-order at low temperature. 

It remains to be seen how the itinerant metamagnetic transition in NiTa$_{4}$Se$_{8}$ contrasts with that in URhGe or UTe$_{2}$, where such a transition is associated with re-entrant superconductivity presumably due to the proliferation of magnetic fluctuations that enhance the pairing susceptibility. Given the apparent proximity of NiTa$_{4}$Se$_{8}$ to superconductivity, and the overall similarity in the magnetic phase diagram between NiTa$_{4}$Se$_{8}$ and URhGe, it would be worthwhile to explore the potential emergence of superconductivity at lower temperatures near the metamagnetic transition in NiTa$_{4}$Se$_{8}$. This may prove useful in linking the physics of heavy fermions to that of the newly discovered unconventional superconductors in layered Van der Waals materials~\cite{kang_cascades_2021,kumar_gate-tunable_2022}.

\section{Conclusion}

In this work, we have presented the synthesis and characterization of a new magnetically intercalated transition metal dichalcogenide, NiTa$_{4}$Se$_{8}$. The system is best described as an itinerant ferromagnet, with spin polarized electronic states in both the local Ni 3$d$ moments and the more itinerant Ta 4$d$ states. By diluting the Ni concentration by a factor of roughly four, superconductivity is observed. The proximity of the ferromagnetic and superconducting phases, together with the unusual spin-polarized Fermi surface topology, is suggestive of ferromagnetic fluctuations as a plausible superconducting pairing mechanism. NiTa$_{4}$Se$_{8}$ is likely not unique among the magnetic TMDs, and it may prove fruitful to look for correlated itinerant magnetism and unconventional superconductivity in TaSe$_{2}$ doped with other magnetic intercalants.

\section{acknowledgements}
This work was supported by QSA, funded by the U.S. Department of Energy, Office of Science, National Quantum Information Science Research Centers. A portion of this work was performed at the National High Magnetic Field Laboratory, which is supported by the National Science Foundation Cooperative Agreement No. DMR-1644779 and the State of Florida. We thank Nobumichi Tamura for assistance with microdiffraction measurements. This research used resources of the Advanced Light Source, which is a DOE Office of Science User Facility under contract no. DE-AC02-05CH11231.
\pagebreak
\bibliography{references}

\pagebreak
\widetext
\begin{center}
\textbf{\large Supplemental Materials: Strongly correlated itinerant magnetism on the boundary of superconductivity in a magnetic transition metal dichalcogenide}
\end{center}
\setcounter{equation}{0}
\setcounter{figure}{0}
\setcounter{table}{0}
\setcounter{page}{1}
\makeatletter
\renewcommand{\theequation}{S\arabic{equation}}
\renewcommand{\thefigure}{S\arabic{figure}}
\renewcommand{\bibnumfmt}[1]{[#1]}
\renewcommand{\citenumfont}[1]{SM#1}

\section*{Methods}
Resistivity and Hall effect measurements were performed in a QuantumDesign PPMS. Gold leads were attached to a crystal using silver paste to a sample cleaved to about 20$\mu$m thickness. A current of 2 mA and 277 Hz frequency was sent through the leads of the sample, and the lognitudinal and Hall voltages were demodulated using a lock-in amplifier. Magnetization and AC susceptibility measurements were performed in a QuantumDesign MPMS and PPMS respectively using a vibrating sample magnetometer. Heat capacity measurements were performed in a QuantumDesign PPMS using a calibrated calorimetry platform with a heater and thermometer. A sample was glued to the platform using Apiezon N grease, the heat capacity of which was measured beforehand. The temperature of the platform was measured while a 2\% temperature rise heat pulse was applied. The sample heat capacity was extracted by fitting the temperature versus time trace during and after the heat pulse. Multiple measurements were taken at each temperature and averaged.

ARPES experiments were conducted at the Beamline 7.0.2 (MAESTRO) at the Advanced Light Source. The data were acquired using the micro-ARPES end station. Samples were cleaved $in-situ$ at the temperature of $\sim$\,150\,K by carefully knocking off alumina posts that are attached on top of each sample with silver epoxy. After the cleaving, the samples were cooled down to the base temperature of $\sim$\,11\,K for measurement under ultra-high vacuum (UHV) better than 4~$\times$~10$^{-11}$ torr. Data were collected with photon energies of 119\,eV. The beam size was $\sim$~15$~\mu$m~$\times$~15$~\mu$m.

Density functional theory calculations were performed using the Vienna Ab Initio Simulation Package (VASP)\citeSM{Kresse1993,Kresse1996}. Projector-augmented wave (PAW)-type pseudopotentials were used with the SCAN meta-GGA exchange correlation functional\citeSM{Sun2015} and spin-orbit coupling was included. The Kohn-Sham equations were solved self-consistently over a 10$\times$10$\times$6 k-point mesh of the Brillouin zone for a plane-wave basis with cutoff energy of 600 eV. As the experimental unit cell differed from the full force-relaxed unit cell by only a marginal amount, we elected to utilize the experimental structure in all calculations shown here to leverage the higher-symmetry of experimentally refined structure. In both cases, the converged magnetic texture remains the same.

\section*{Data availability statement}
All data and analysis code provided in this report are publicly available: DOI 10.17605/OSF.IO/27RJE.

\pagebreak
\section*{Additional magnetization measurements of $T^{*}$}

\begin{figure}[!htbp]
\centering
\includegraphics[scale=0.7]{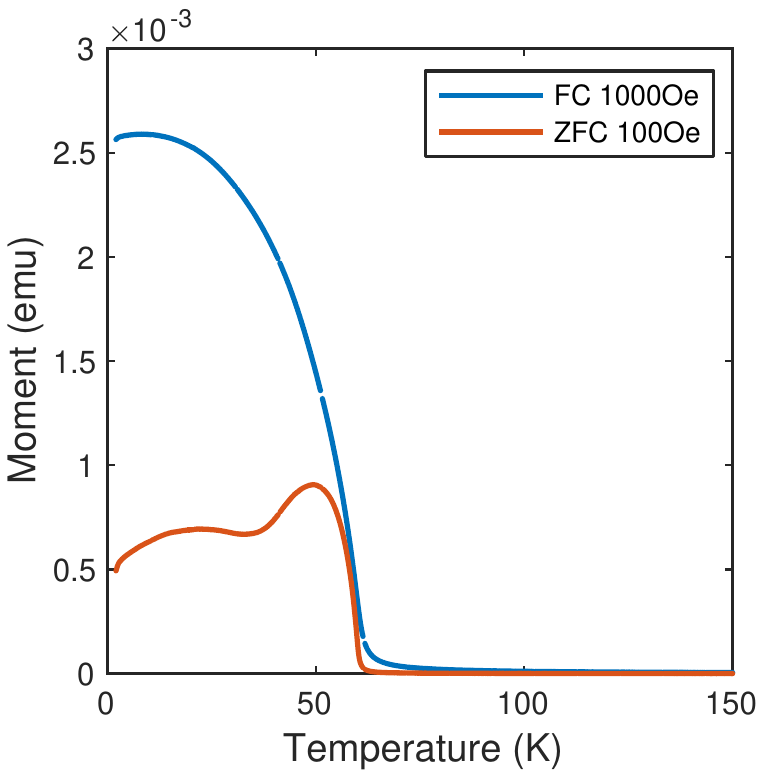}
\caption{{\bf Magnetization data on polycrystalline NiTa$_{4}$Se$_{8}$} Field-cooled (FC) and zero field-cooled (ZFC) protocols. A main ferromagnetic-like transition is observed at 58K, and a secondary anomaly is observed around 36K in qualitative agreement with the data on single crystal samples.}
\label{fig:sup_powdermag}
\end{figure}

This section focuses on additional measurements of the crossover scale $T^{*}$, which is primarily explored in Fig.~\ref{fig:phases} of the main text. First, Fig.~\ref{fig:sup_powdermag} presents magnetization data taken on a powder sample with two different field cooling protocols. The powder measured here is obtained from the prereacted materials before they are put through the chemical vapor transport process (see also the crystal growth section of the main text). The overall trend is similar to that observed on single crystal samples. The shape of the curve is qualitatively different from the data in Figs.~\ref{fig:phases}A,B likely because the data on powder averages the susceptibility over all field orientations. Nevertheless, the powder data exhibits a ferromagnetic-like transition at $T_{C} = 58$K, and a secondary feature at $T^{*} = 36$K.

\section*{Hall effect measurements on a single crystal sample}

\begin{figure}[!htbp]
\centering
\includegraphics[scale=0.7]{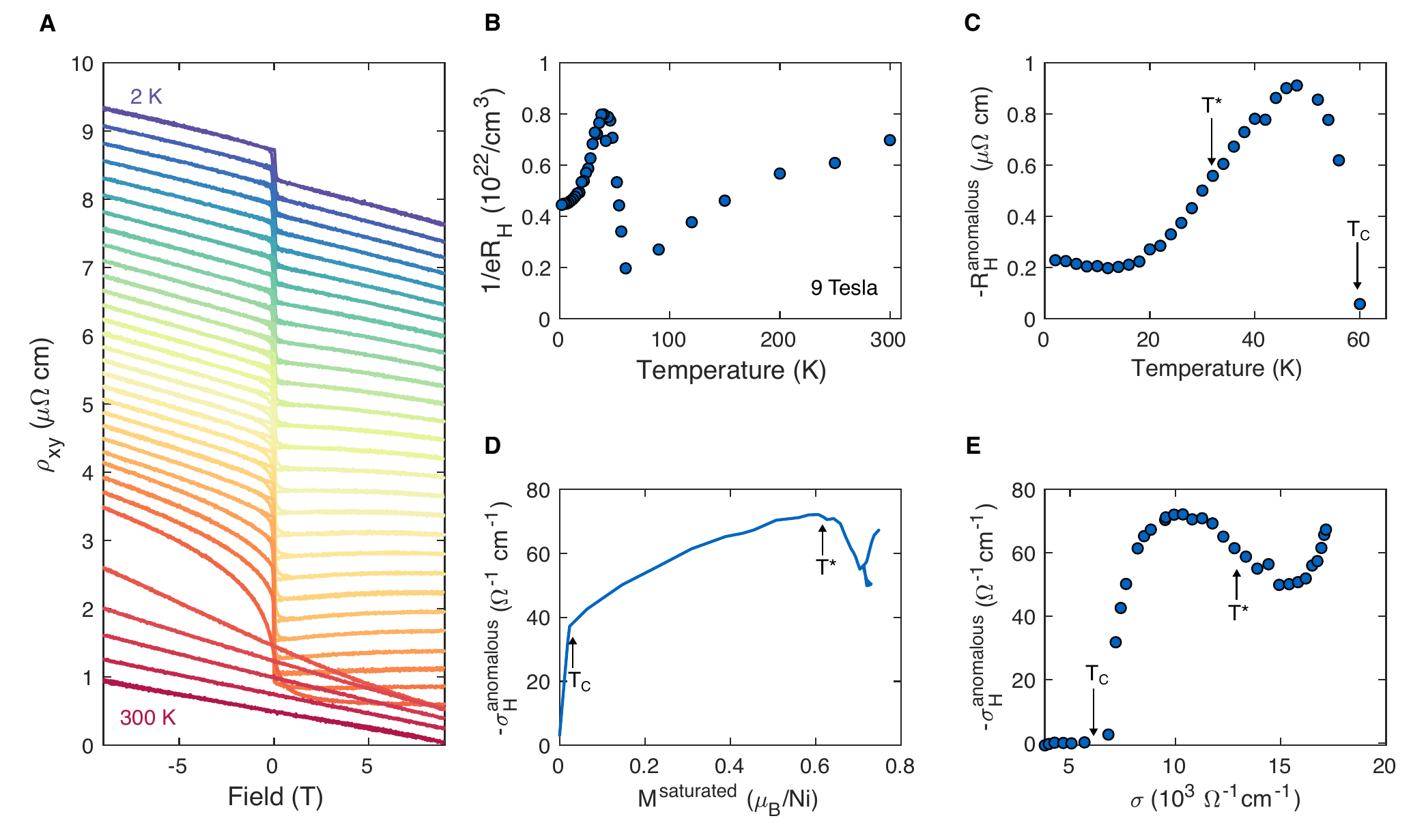}
\caption{{\bf Hall effect data on single crystal sample with field directed out of the plane}~\textbf{A} Positive and negative field sweeps of the Hall resistivity ($\rho_{xy}$) at different temperatures.~\textbf{B} Inverse of the conventional Hall coefficient ($R_{H}$) times the electron charge, extracted from the high field value of $\rho_{xy}/B$ at a field of 9 Tesla after subtracting the positive and negative field components of the traces shown in panel A.~\textbf{C} Anomalous Hall coefficient, defined as the difference in $\rho_{xy}$ at zero field between positive and negative field sweeps divided by 2.~\textbf{D} Anomalous Hall conductivity as a function of the saturated magnetization.~\textbf{E} Anomalous Hall conductivity as a function of the longitudinal conductivity.}
\label{fig:sup_Hall}
\end{figure}

Fig.~\ref{fig:sup_Hall}A shows the results of Hall effect measurements on a single crystal sample of about 20$\mu$m thickness. The field was swept from positive 9 Tesla to negative 9 Tesla and back. A prounounced curvature in the low-field Hall resistivity develops upon decreasing temperature. Below the Curie temperature, an anomalous Hall component develops on top of the conventional Hall effect with overall electron-like character.

There are several ways to analyze anomalous Hall effect data, but generally speaking the Hall effect in ferro- or ferrimagnets has the following phenomenological form
\begin{equation}
    \rho_{xy} = R_{H}(B) + R_{H}^{anomalous}(M).
\end{equation}
The first term, $R_{H}$(B), represents the standard Hall coefficient coming from the action of the Lorentz force on charged carriers, which is controlled by the mobility and density of the different carrier densities present in the system and depends on external magnetic field, $B$, in a generally complicated way. The second term represents the anomalous component stemming from the spontaneous magnetization in the sample, $M$. 

In Fig.~\ref{fig:sup_Hall}B, we plot the value of the conventional Hall coefficient obtained via value of $\rho_{xy}(B)/B$ at 9 Tesla where, in order to account for misalignment in the electrodes and remove the anomalous Hall contribution, the value is anti-symmetrized between positive and negative field ($R_{H}(9T) = (\rho_{xy}(9T) - \rho_{xy}(-9T))/(2 \times 9T)$). In a system with one species of electrical charge carrier, this value is inversely proportional to the net carrier density. However, in this material, it is likely that there are both electron and hole carriers given the calculated Fermi surfaces shown in Fig.~\ref{fig:band}. In such a scenario, the high-field limiting value of $R_{H}$ is inversely proportional to the sum of all carrier densities of each different species. However, this limit is only reached where $\rho_{xy}$ exhibits an extended linear-in-field dependence at high fields. There are several temperatures where this criterion is not met --- for example, immediately above the ferromagnetic transition the Hall resistivity is markedly nonlinear. Interestingly, however, it appears that across both $T_{C} = 58K$ and at $T^{*} = 36K$, the value of $1/eR_{H}$ at 9 Tesla exhibits notable features. At $T_{C}$, the inverse Hall coefficient steeply rises, and then decreases again below $T^{*}$. It may be possible to interpret $R_{H}$ in terms of the net carrier density at these temperatures given that $\rho_{xy}$ is quite linear in field over this whole range. In this scenario, $T^{*}$ could be associated with a change in carrier density of the system, which could happen if some carriers are gapped by a magnetic exchange interaction. However, at present it is not possible to rule out the possibility that the high-field limit has not been reached, and that the carrier mobility depends strongly on temperature across this transition --- a reasonable possibility given that the longitudinal resistance also exhibits a change in curvature across $T^{*}$. Further measurements at higher fields are required to solidify the interpretation of the Hall coefficient between $T_{C}$ and $T^{*}$; nevertheless, the direct extraction of the conventional Hall coefficient as a function of temperature seems to show signatures of qualitative changes across $T^{*}$.

The anomalous Hall resistivity ($R_{H}^{anomalous}$) generally depends on the spontaneous magnetization of the sample ($M$), as well as potentially the scattering rate and carrier density, which are typically accounted for by comparing $R_{H}^{anomalous}$ to the longitudinal resistivity or conductivity. Fig.~\ref{fig:sup_Hall}C shows the value of the anomalous Hall contribution to the Hall resistivity as a function of temperature. The value steeply rises below $T_{C}$, and then gradually decreases with temperature --- such behavior is fairly typical of ferromagnets. From this data, the anomalous Hall conductivity can be calculated using the following formula:
\begin{equation}
    \sigma_{H}^{anomalous} = \frac{\rho_{xy}^{anomalous}}{\rho_{xx}^{2} + (\rho_{xy}^{anomalous})^{2}},
\end{equation}
where $\rho_{xx}$ is the longitudinal resistivity. Typical analysis of the anomalous Hall component involves plotting the anomalous Hall conductivity against the longitudinal conductivity, as shown in Fig.~\ref{fig:sup_Hall}E~\citeSM{nagaosa2010anomalous}. The anomalous Hall conductivity could have contributions from intrinsic sources, such as a contribution from the Berry phase of the band structure, or extrinsic sources, such as skew scattering. Intrinsic sources typically lead to $\sigma_{H}^{anomalous} \sim \sigma^{0}$ for moderately dirty samples or $\sigma_{H}^{anomalous} \sim \sigma^{1.6}$ in very dirty samples. For extrinsic sources of anomalous Hall conductivity, $\sigma_{H}^{anomalous}$ is proportional to $\sigma$ in a non-universal way, but the two should still have a positive correlation. We see that in Fig.~\ref{fig:sup_Hall}E, the anomalous Hall conductivity is mostly independent of the longitudinal conductivity, but exhibits a rather unusual dependence across $T^{*}$ in that $\sigma_{H}^{anomalous}$ actually decreases with $\sigma$ before increasing again at low temperature. Therefore, overall the anomalous Hall effect in this material seems to have an intrinsic origin over a wide range of temperature, but there seems to be a qualitative change in the transport mechanism across $T^{*}$, which leads to a moderate suppression of the anomalous Hall conductivity. In the conventional analysis described above, of course the carrier density is assumed to be constant. Thus, the atypical behavior observed in Fig.~\ref{fig:sup_Hall}E could originate from the possibly temperature-dependent carrier density as suggested by the conventional Hall contribution discussed in the previous paragraph. However, there may be more complicated temperature-dependent electronic structure or mobility effects across $T^{*}$ which produce the subtle negative correlation in Fig.~\ref{fig:sup_Hall}E. A similar phenomenon is observed in Fig.~\ref{fig:sup_Hall}D, where we plot the anomalous Hall conductivity as a function of the saturated ferromagnetic moment ($M^{saturated}$). There is an expected positive correlation between the two up until $T^{*}$, at which point the anomalous Hall conductivity actually decreases with the size of the saturated moment. Such behavior is atypical of anomalous Hall ferromagnets. Again, this seems to indicate a qualitative change in the anomalous Hall transport mechanism across $T^{*}$. Finally, we note that the overall the size of the anomalous Hall conductivity ($\sim 70 \Omega^{-1}cm^{-1}$) is fairly modest in this material relative to other anomalous Hall ferromagnets, as is the size of the anomalous Hall angle ($\sigma_{H}^{anomalous}/\sigma \leq 1\%$). 

\pagebreak
\section*{Bipartite structure in $c$-axis hysteresis loops}

\begin{figure*}[!htbp]
\centering
\includegraphics[scale=0.7]{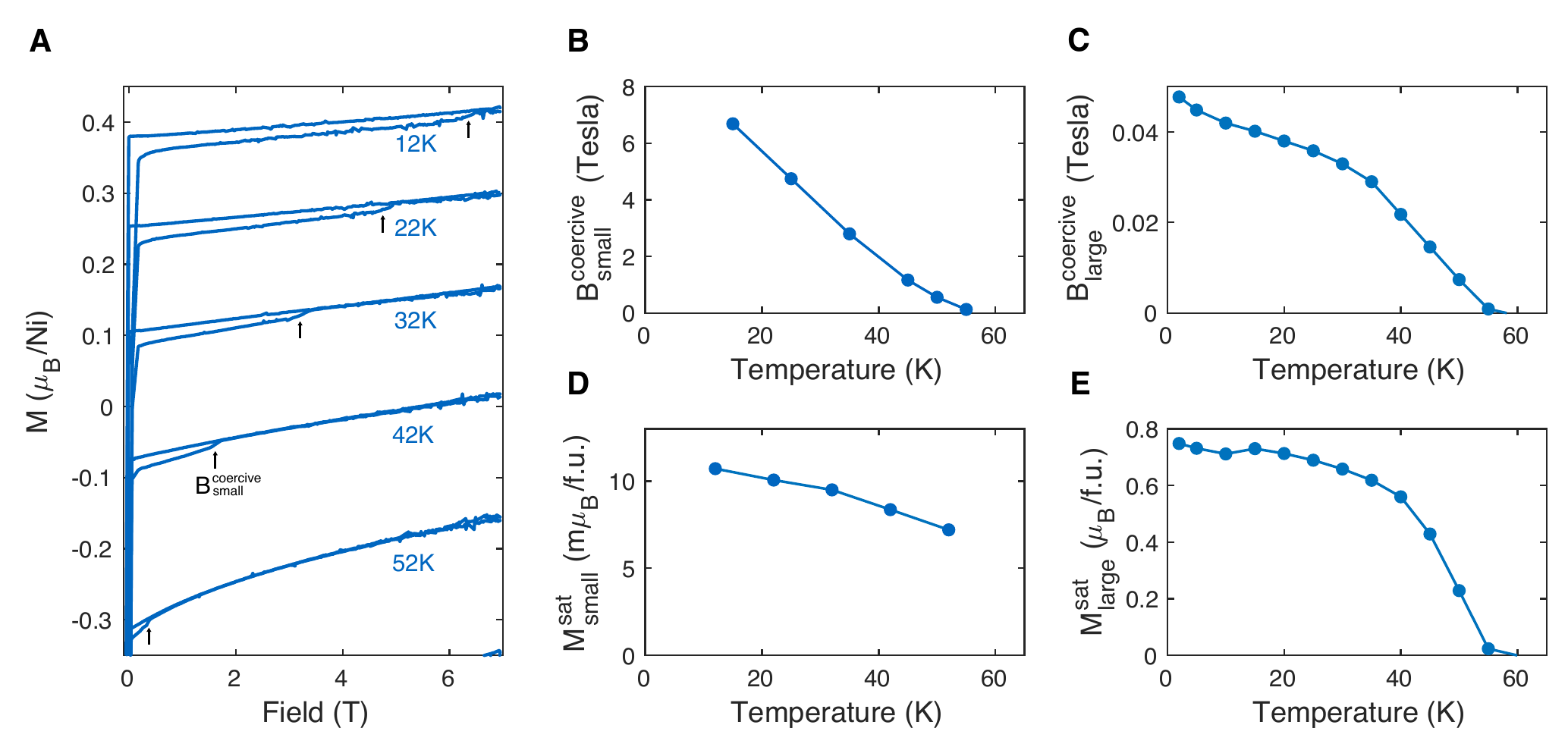}
\caption{{\bf Coercive field and saturated moment measurements for magnetic field in the interplanar direction}~\textbf{A} Magnetization versus field hysteresis loops zoomed in on the positive field region. The curves are offset vertically for clarity. The hysteresis loop has two components. On top of the dominant hysteresis loop associated with a large moment, a coercive field event associated with a small moment (black arrows) is observed at relatively higher fields.~\textbf{B} Temperature-dependence of the coercive field of the small moment.~\textbf{C} Temperature-dependence of the coercive field of the small moment.~\textbf{D} Temperature-dependence of the small saturated moment.~\textbf{E} Temperature-dependence of the large saturated moment.}
\label{fig:sup_coercive}
\end{figure*}

Fig.~\ref{fig:sup_coercive}A shows magnetization versus field traces for field applied along the $c$-axis of the crystal. A main coercive event is observed centered around zero field with a relatively large saturated moment. At higher fields, a secondary coercive field event is observed with a relatively much smaller saturated moment, as delineated in the figure by black arrows. The temperature dependence of the coercive field of both events, and the associated jump in sample magnetization, are shown in Figs.~\ref{fig:sup_coercive}B-E.

The coercive field of the large jump in magnetization exhibits a somewhat non-trivial dependence on temperature (Fig.~\ref{fig:sup_coercive}C). While it does rise with decreasing temperature below $T_{C}$, the ferromagnetic transition, it exhibits an inflection point around $T^{*} = 36$K. The coercive field is generally understood to be set by the single-ion anisotropy, and the pinning strength of domains within the sample. It is possible that $T^{*}$ is associated with some sort of change in the domain structure of the dominant ferromagnetic moments, resulting in a change in the temperature-dependent coercive field. The coercive field of the small jump in magnetization rises steeply below $T_{C}$ (Fig.~\ref{fig:sup_coercive}B). The magnetization jump of both the small event and large event rise with decreasing temperature (Figs.~\ref{fig:sup_coercive}D,E). This trend is more pronounced for the large magnetization jump. 

\pagebreak
\section*{Metamagnetic transition}

\begin{figure}[!htbp]
\centering
\includegraphics[scale=0.7]{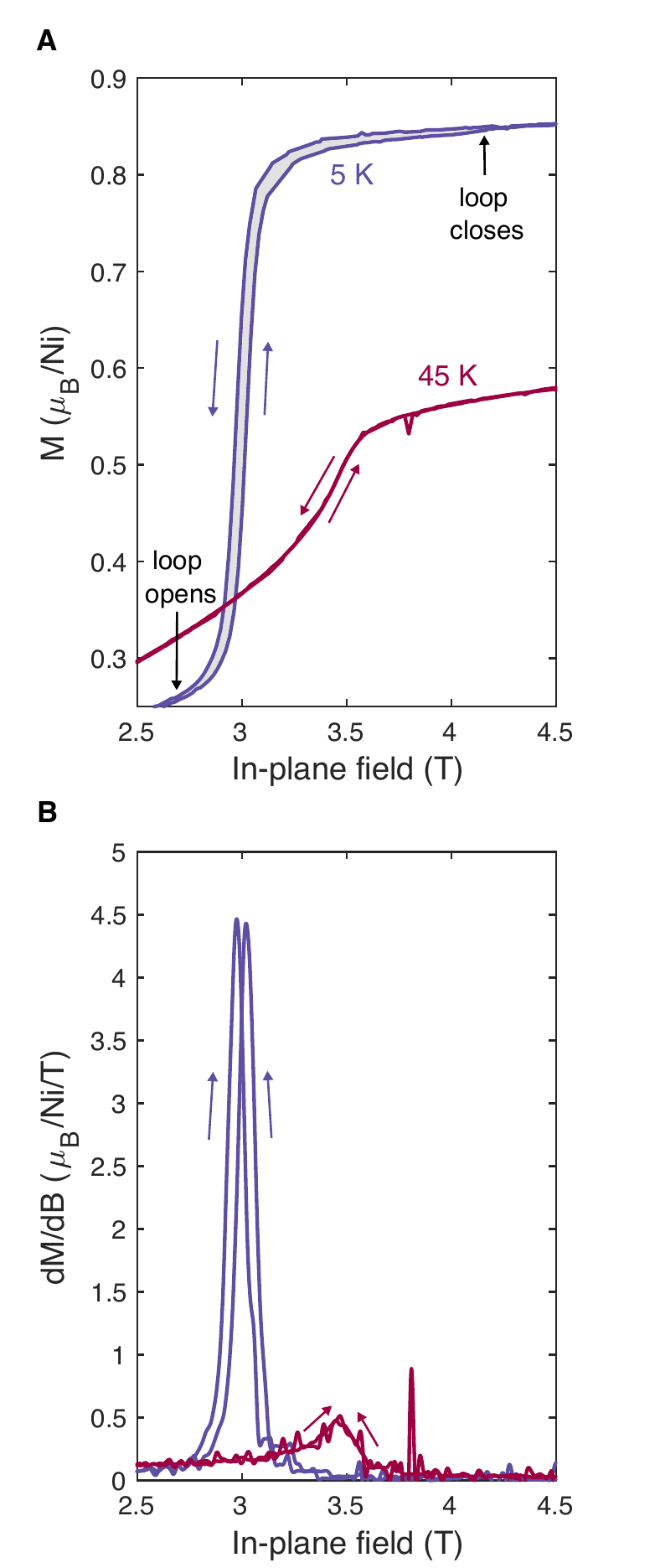}
\caption{{\bf Magnetization sweeps of the in-plane metamagnetic transition}~\textbf{A} Magnetization versus field at 5K and 45K. At low temperature, the transition is associated with a hysteresis loop shaded in grey. At higher temperatures, the up and down sweeps are indistinguishable suggesting that the transition is continuous.~\textbf{B} Derivative of magnetization with respect to field for the same data in panel A.}
\label{fig:sup_inplanetransition}
\end{figure}

In this section we further explore the behavior of the metamagnetic transition observed when field is directed along the intraplanar direction of single crystals as seen in the main text in Fig.~\ref{fig:magnetization}B. First, we introduce the hysteretic behavior of the metamagnetic transition as a function of temperature. Fig.~\ref{fig:sup_inplanetransition}A shows the magnetization versus field traces zoomed in on the metamagnetic transition. Hysteresis in this transition is present at 5K but not at 45K. This behavior is also seen in the derivative of the magnetization (Fig.~\ref{fig:sup_inplanetransition}B). Conventionally, the presence of hysteresis in a magnetic transition is taken as evidence that the transition is first-order, while second-order transitions on general grounds do not exhibit hysteresis. Note also that the metamagnetic transition manifests as a peak in $dM/dB$.

\begin{figure}[!htbp]
\centering
\includegraphics[scale=0.65]{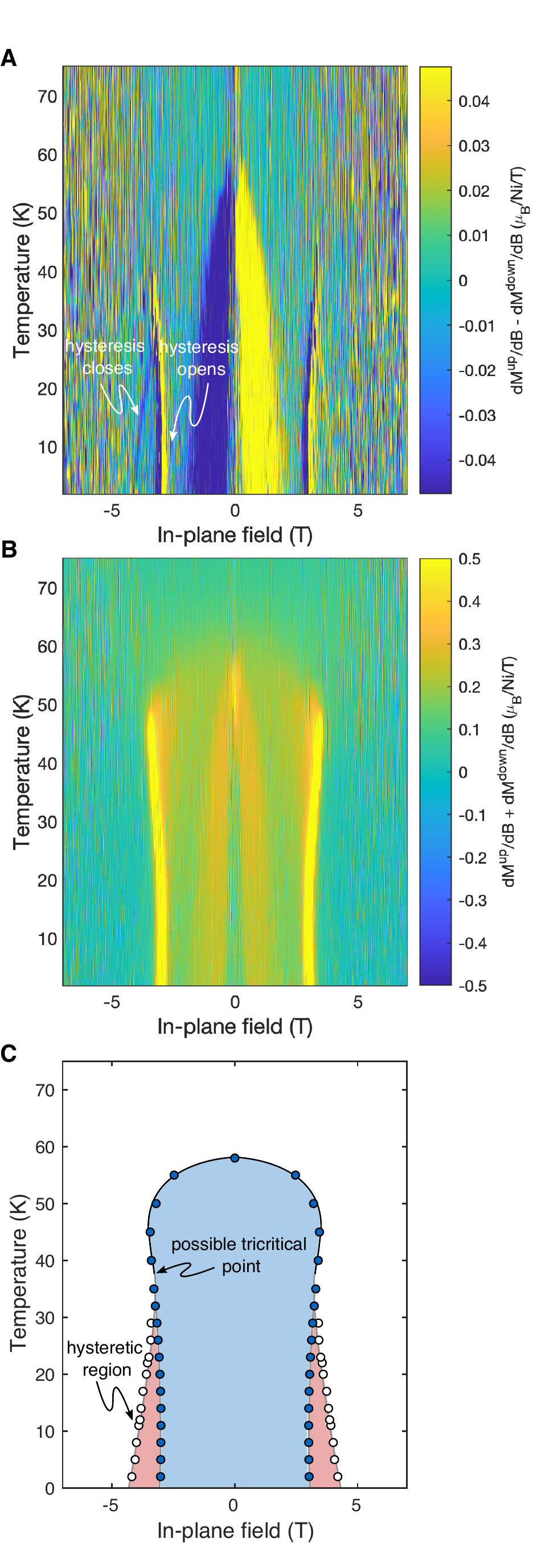}
\caption{{\bf Phase diagram constructed with susceptibility measurements for in-plane magnetic field}~\textbf{A} Derivative of magnetization with respect to field, $dM/dB$. The difference between up and down magnetic field sweeps is used to resolve the two edges of the hysteresis loop. The metamagnetic transition becomes hysteretic below about 30K, where the start and end of the hysteresis loop are observed as separate bright features in this color map.~\textbf{B} Average of up and down sweeps of $dM/dB$. The centroid of the metamagnetic transition is resolved as a bright band in this color map.~\textbf{C} Phase diagram constructed with blue points delineated peaks determined from panel B. White points delineate the boundaries of the hysteretic region around the metamagnetic transition determined from panel A. The phase boundary line is drawn as black if it is not hysteretic (no feature in panel A), and grey if it is hysteretic (features in panel A). A possible multicritical point where the transition changes from non-hysteretic to hysteretic is indicated.}
\label{fig:sup_phasediagrams}
\end{figure}

Fig.~\ref{fig:sup_phasediagrams} focuses on tracking the transition field and the presence of hysteresis as a function of temperature. We choose to track the transition field by averaging the up and down sweeps of $dM/dB$ (Fig.~\ref{fig:sup_phasediagrams}B). In this color plot, the metamagnetic transition appears as a bright band. The features near zero field are signatures of a hysteresis loop centered at zero field, which are distinct from the metamagnetic transition. This zero-field hysteretic behavior is explored more in the section ``Free energy models of magnetization loops''.

As seen in Fig.~\ref{fig:sup_inplanetransition}A, the metamagnetic transition is hysteretic at low temperature, but non-hysteretic at higher temperature. In order to explore the evolution of this behavior, we present a color plot of the derivative of the magnetization $dM/dB$ where the up and down sweeps of field have been subtracted from each other. Features in such a color plot are associated with differences between the up and down sweeps of magnetic field, i.e. hysteresis. In fact, we do observe two separate bright bands between 3.5 Tesla and 4.5 Tesla in Fig.~\ref{fig:sup_phasediagrams}A. Each bright band is associated with the opening and closing of the hysteresis loop around the metamagnetic transition, as seen in a single trace in Fig.~\ref{fig:sup_inplanetransition}A. The two bands join together and disappear at 40K. This trend indicates that the metamagnetic transition changes from second-order (non-hysteretic) to first-order (hysteretic) upon decreasing temperature below 40K. Such behavior is potentially consistent with the presence of a tricritical point at the point in the phase diagram where the metamagnetic transition changes from first to second order, i.e. at 3.5 Tesla and 40K for field directed in the plane of the sample. This possibility should certainly be explored further.

The overall temperature-dependent phase diagram for in-plane magnetic field is summarized schematically in Fig.~\ref{fig:sup_phasediagrams}C. Blue data points are taken from the peak in the $dM/dB$ curve averaged for up and down sweeps of magnetic field (Fig.~\ref{fig:sup_phasediagrams}B) --- this demarcates the central part of the metamagnetic transition. White points indicate the closing of the hysteresis loop, which is associated with a small feature that is most readily observed if up and down sweeps of $dM/dB$ are subtracted from each other (Fig.~\ref{fig:sup_phasediagrams}A).

\begin{figure}[!htbp]
\centering
\includegraphics[scale=0.7]{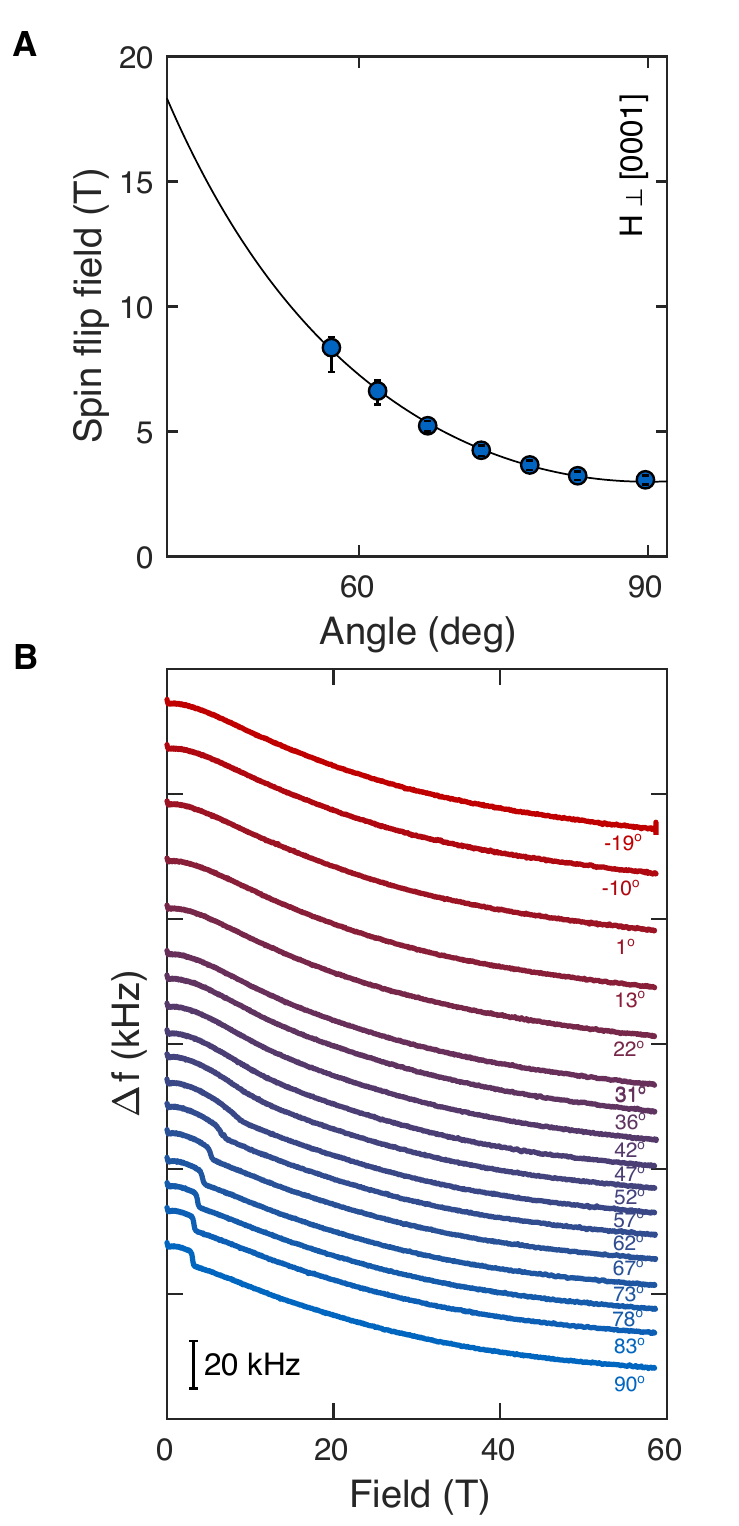}
\caption{{\bf Metamagnetic transition as detected by proximity detector oscillator (PDO) measurements as a function of field-angle tilting from out-of-plane to in-plane directions. The sample temperature is between 1.58 K and 1.61 K}~\textbf{A} Transition field as a function of angle, extracted from derivatives of the field traces shown in panel B. The black line is 24.6($\frac{1.12}{\rm{sin}(\theta)} - 1)$.~\textbf{B} Field sweeps of the resonant frequency shift of a PDO circuit including the sample; curves are offset vertically for clarity. The shift of the resonant frequency is proportional to the electrical conductance of the sample \cite{altarawneh2009proximity}. A transition is observed when the field is in the plane of the crystal, consistent with the magnetization data in Fig.~\ref{fig:magnetization} of the main text. The transition rapidly disperses to higher fields as the angle tilts out-of-plane, and disappears when the tilt angle is lower than about 60 degrees.}
\label{fig:sup_PDO}
\end{figure}

We now move on to exploring the dependence of the metamagnetic transition field on the tilt angle from in-plane to out-of-plane. Because such a transition could in principle lie well beyond the 7 Tesla field range of our vibrating sample magnetometry instrument, we opted to perform measurements up to 60 Tesla in pulsed magnetic fields at the National High Magnetic Field Lab at Los Alamos National Lab. These measurements were carried out using a proximity detector oscillator circuit, which consists of a tank oscillator attached to a small pancake-shaped coil (see also \citeSM{altarawneh2009proximity} for more details). The single crystal of NiTa$_{4}$Se$_{8}$ is glued onto the surface of the pancake coil. Changes to the resonant frequency of this circuit are detected during the course of the field pulse, and are associated primarily with changes in the inductance of the sample coupled to the coil. This essentially measures the skin depth of the sample for metallic samples, which is proportional to the material's conductivity. As seen in Fig.~\ref{fig:sup_PDO}B, the metamagnetic transition is clearly observed as a sharp jump in the resonant frequency of the PDO circuit as the field increases above 3 Tesla for field directed along the basal planes of the crystal (in the language of the figure, this corresponds to 90 degree tilt angle). No further features are observed between 3 Tesla and 60 Tesla --- the overall trend with increasing field above the metamagnetic transition is fairly typical behavior for the magnetoconductivity of a metal. This indicates that the magnetization of the sample is fully polarized above 3 Tesla, and there are no further metamagnetic transitions between 3 and 60 Tesla. As the field is tilted in to the out-of-plane orientation, the transition becomes less sharp and moves to higher field. By taking a derivative of the curve, we are able to extract the metamagnetic transition field as a function of angle, as shown in Fig.~\ref{fig:sup_PDO}A. The angle-dependence of the transition is well-described by an inverse sinusoidal form, indicating that the component of the field in the plane of the crystal is the main driving factor which induces the metamagnetic transition. When the field is tilted less than 60 degrees away from the normal to the basal planes of the crystal, no metamagnetic transition is discernible in the conductance measurements.

\pagebreak
\section*{Free energy models of magnetization loops}

In order to capture some of the features of the magnetization data shown in Fig.\ \ref{fig:magnetization} of the main text, we explore a minimal free energy model for the nickel and tantalum subsystems. To recap, the salient features of the magnetic measurements are (1) two coercive field events when measuring $M$ for $ \mathbf{H} \parallel \mathbf{c}$, (2) a hysteresis loop about 0 T when measuring $M$ for $\mathbf{H} \perp \mathbf{c}$, and (3) a metamagnetic transition in the same configuration. As explained above, the metamagnetic transition is likely a band structure effect unique to itinerant magnets. Here, we focus on remaining two features of the data.

\begin{figure}
    \centering
    \includegraphics{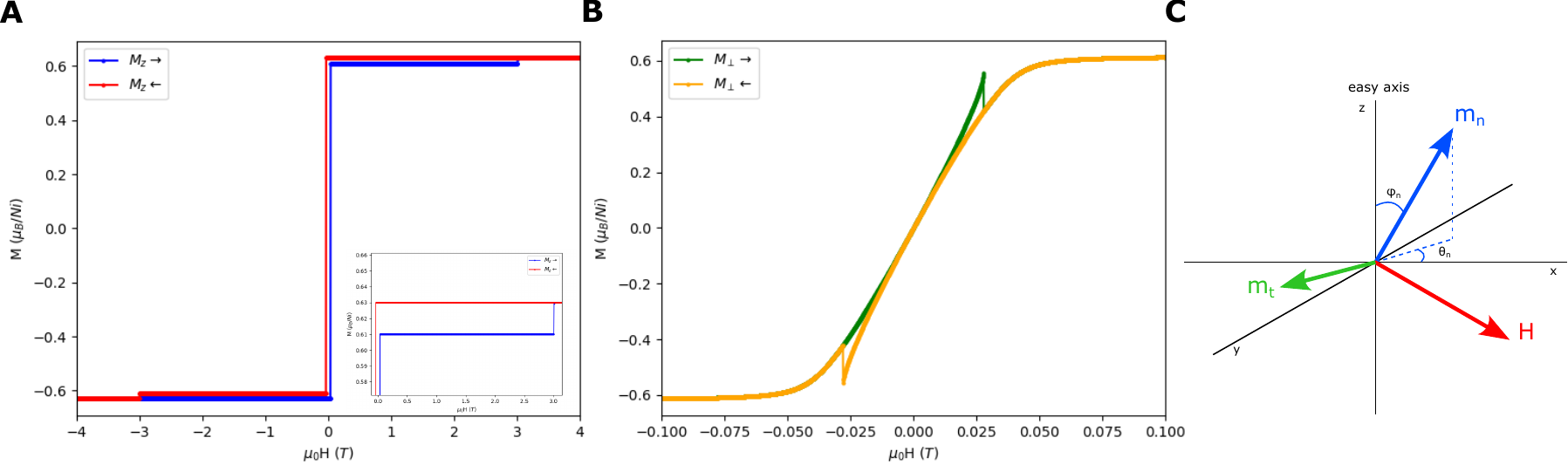}
    \caption{Simulated magnetization loops according to Eqn.\ \ref{eq:energy} for $J=0$. Parameters have been tuned to best match the experiment at 35 K. \textbf{A} Applied field out-of-plane, and  \textbf{B} applied field in-plane. \textbf{C} Definition of coordinate systems. This non-interacting model can capture the bipartite structure in magnetization out-of-plane, but fails to describe the in-plane hysteresis.}
    \label{fig:sup_free_energy_uncoupled}
\end{figure}


The system is described by two order parameters, one corresponding to the nickel moment $m_n$ and the other to the tantalum moment $m_t$. Assuming a single magnetic domain, the macroscopic free energy can be written to first order consistent with hexagonal symmetry as:

\begin{equation}
\begin{aligned}
    E(\phi_n, \phi_t, \alpha, h) = &K_n m_n^2 \sin\phi_n^2 + K_t m_t^2 \sin\phi_t^2 \\
    &- \mu_0 h \big( m_n \cos(\alpha - \phi_n) + m_t \cos(\alpha - \phi_t) \big) \\
    &- J \cos(\phi_n - \phi_t) \\
    \label{eq:energy}
\end{aligned}
\end{equation}

\noindent where $K_n$ and $K_t$ are the single-ion anisotropy for the nickel and tantalum atoms respectively, $J$ is the exchange coupling between nickel and tantalum systems, $\phi_n$ and $\phi_t$ are the angles between nickel and tantalum moments to the out-of-plane axis, respectively, $\alpha$ is the angle between the field and the same axis, and $h$ is the field strength. 

We obtain hysteresis loops according to this model by solving numerically via a gradient descent method. For a fixed field angle $\theta$, an initial configuration for $\phi_n$ and $\phi_t$ can be advanced as $h$ is swept from positive to negative (and back) by, at each step, relaxing the system to a local minimum according to the gradient of Eqn.\ (\ref{eq:energy}). As illustrated in Fig.\ \ref{fig:sup_free_energy_uncoupled} and described below, we are able to reproduce the bipartite structure in out-of-plane magnetization from this model.

To obtain reasonable parameters for the constants, we first choose units such that $\mu_0 = 1$ and set $m_n = 0.62$ and $m_t = 0.01$ in units of $\mu_B/$Ni, as determined from the data and density functional calculations. We then tune $K_n$ and $K_t$ in the uncoupled case ($J=0$) until the first and second coercive field events match the data at $T = 35$ K. This iterative procedure yields optimal anisotropy constants $K_n = 0.028$ and $K_t = 150$. Since anisotropy originates from a spin-orbit matrix element, these constants can be regarded as an intrinsic strength of the spin-orbit coupling. For the anisotropy energy of these two subsystems to be comparable, there must be a much greater intrinsic strength for the small tantalum moment in comparison to the larger nickel moment --- this is consistent with the fact that the spin-orbit interaction is on general grounds stronger in 4$d$ and 5$d$ electrons than in 3$d$. The temperature dependence of the second coercive field event could be caused by a combination of temperature induced changes in the tantalum spin-orbit coupling and the acquired moment.

\begin{figure}
    \centering
    \includegraphics{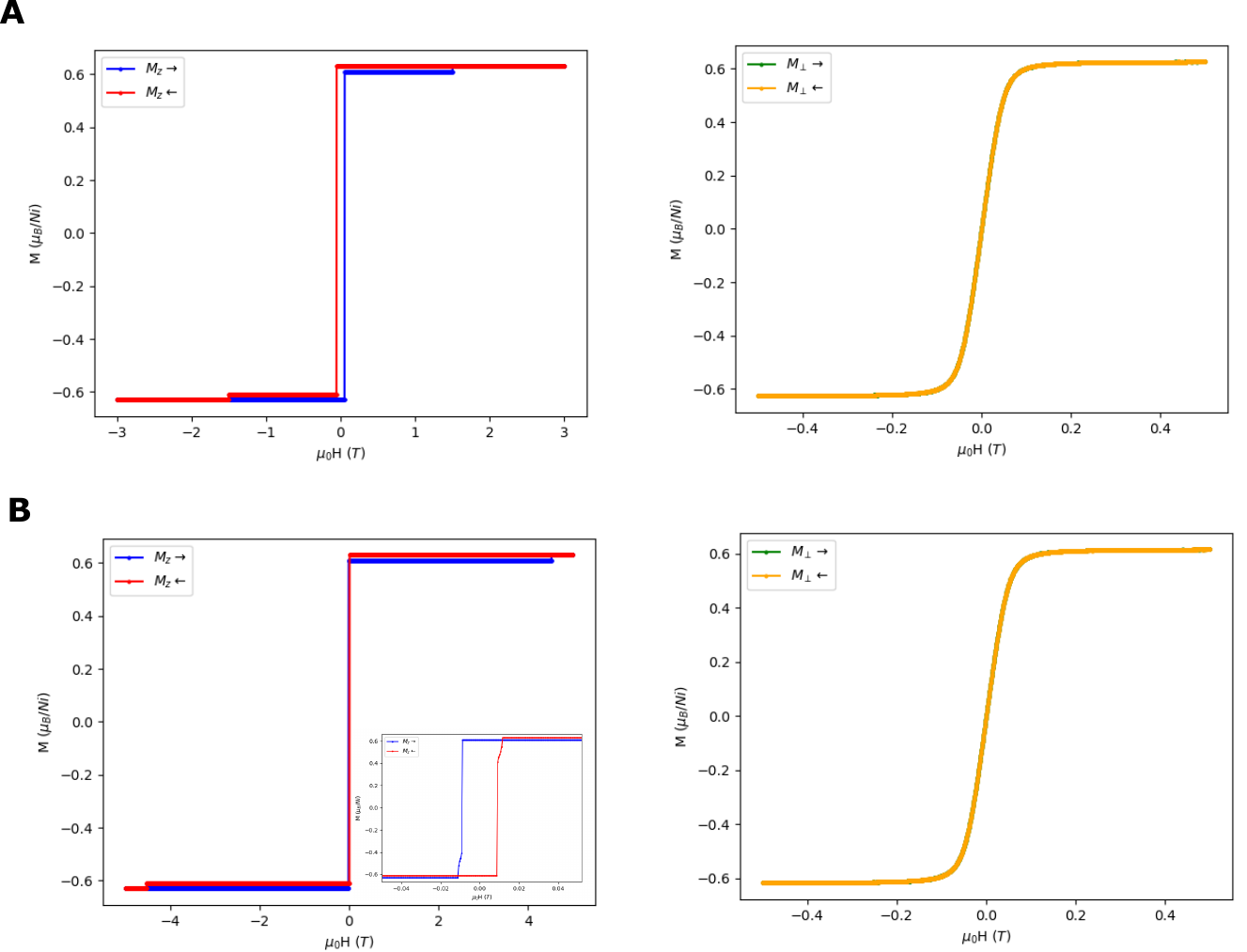}
    \caption{Simulated magnetization loops in the coupled nickel and tantalum system for \textbf{A} ferromagnetic with $J = K_tm_t^2$, and \textbf{B} antiferromagnetic with $J = -K_tm_t^2$. Out-of-plane, ferromagnetic coupling suppresses the bipartite structure, whereas antiferromagnetic coupling enhances it and inverts the hysteresis loop. In-plane, both curves are qualitatively similar to the uncoupled case.}
    \label{fig:sup_free_energy_coupled}
\end{figure}

The addition of ferromagnetic exchange ($J>0$) strongly suppresses the bipartite structure along the out-of-plane axis, as shown in Fig.\ \ref{fig:sup_free_energy_coupled}A. This can be understood by considering $J$ to be very large, in which case the nickel and tantalum moments will track together as the field is swept. Clearly, there must be some crossover when $J \sim K_t m_t^2$. The presence of a bipartite structure thus indicates that the two subsystems must be weakly coupled. The effect of antiferromagnetic exchange ($J<0$) is shown in Fig.\ \ref{fig:sup_free_energy_coupled}B.

While the model in Eqn.\ \ref{eq:energy} can reproduce the bipartite structure in the out-of-plane direction, it fails to capture the in-plane hysteresis. In fact, given the small tantalum moments and the energy scales determined above, it seems unlikely that such a large hysteresis loop can be related to the tantalum system. The simplest explanation for such an effect is biaxial anisotropy of the nickel moments.

\begin{figure}
    \centering
    \includegraphics{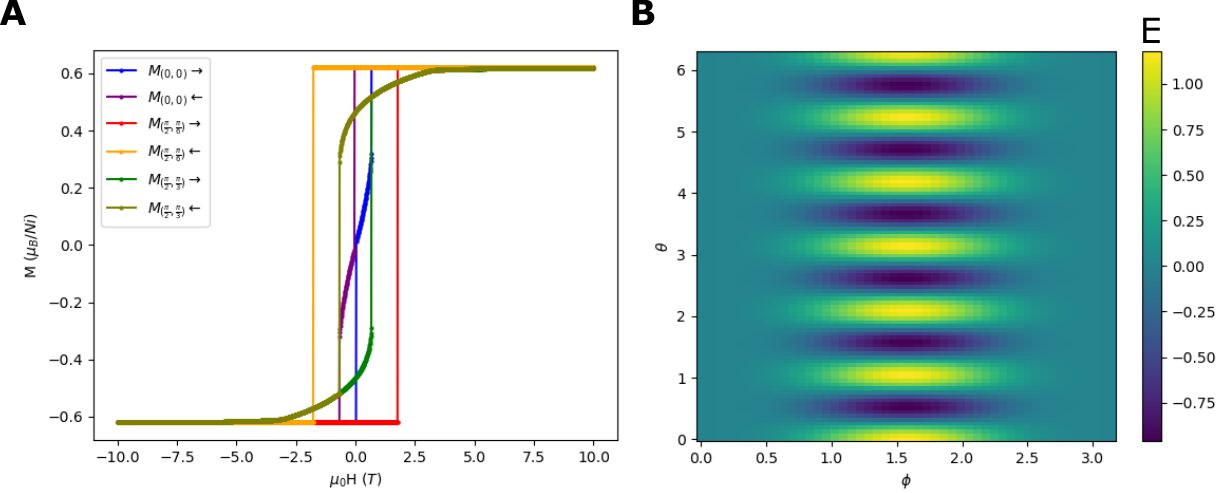}
    \caption{\textbf{A} Simulated magnetization loops according to Eqn.\ \ref{eq:energy2} for $K_1 = K_n$, $K_2 = 0$, and $K_3 = 10K_1$, illustrating how an enlarged in-plane hysteresis loop corresponds to a switching of the easy axis for hexagonal crystals. For each curve, a field is applied along $(\alpha, \beta)$ and the magnetization is measured in the same direction. \textbf{B} The anisotropy energy at $H=0$ showing and energy minimum in the plane.}
    \label{fig:sup_free_energy_biaxial}
\end{figure}

\begin{figure}[!htbp]
\centering
\includegraphics[scale=0.7]{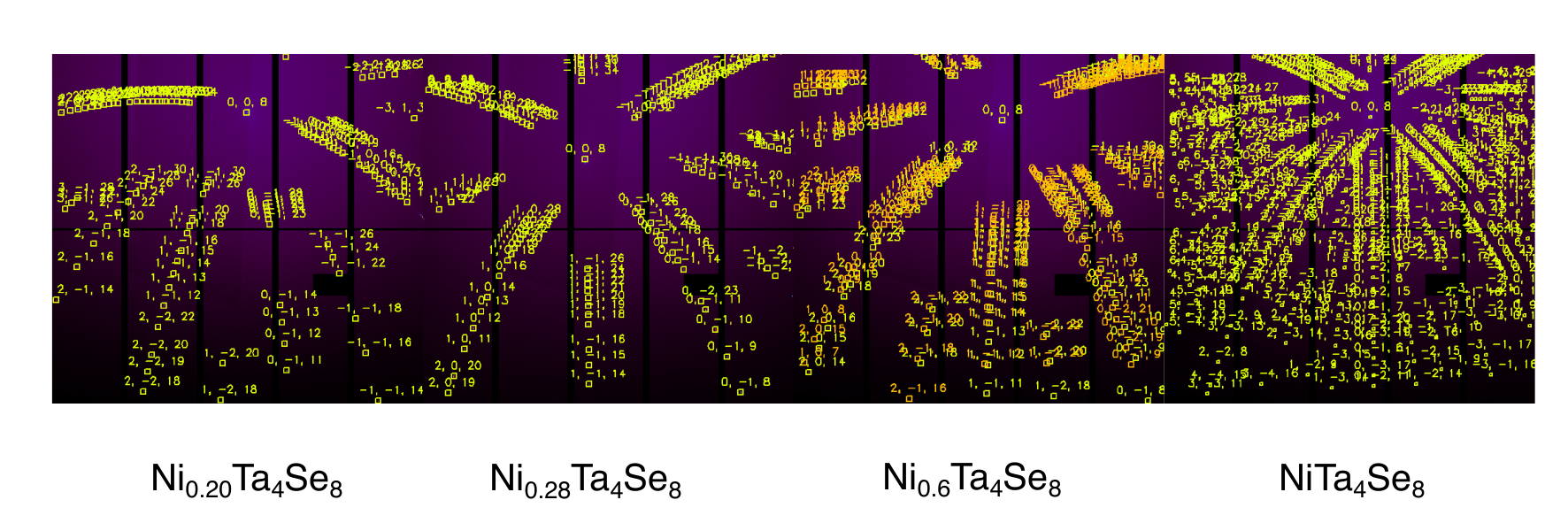}
\caption{{\bf Microlaue x-ray diffraction measurements conducted on Ni$_{x}$Ta$_{4}$Se$_{8}$ samples} The spot size in the measurements is on the order of several micrometers in a Laue scattering geometry. NiTa$_{4}$Se$_{8}$ samples are characterized by twice as many peaks as the pure TaSe$_{2}$ structure as a result of doubling of the unit cell corresponding to the superstructure of the triangular lattice of nickel atoms. These doubled peaks are absent in samples with between 0.2 and 0.28 nickel concentration, suggesting that the nickel atoms form a disordered layer between the TaSe$_{2}$ sheets (the reflections associated with the TaSe$_{2}$ layers are still present). In the sample with 0.6 nickel concentration, the sharp peaks associated with the superstructure are not present, and there appear to be two separate sets of peaks indicating a high degree of twinning within the structure of the TaSe$_{2}$ layers.}
\label{fig:sup_dopedstructures}
\end{figure}

To study biaxial anisotropy, we consider the nickel subsystem alone and expand the free energy to sixth order in the directional cosines, at which order the in-plane symmetry is broken,

\begin{equation}
\begin{aligned}
    E(\phi, \theta, \alpha, \beta, h) = &K_1 m_n^2 \sin\phi^2 + K_2 m_n^2 \sin\phi^4 + K_3 m_n^2 \sin\phi^4\cos6\theta \\
    &- \mu_0 h m_n \big(\sin\phi\cos\theta\sin\alpha\cos\beta + \sin\phi\sin\theta\sin\alpha\sin\beta \\
    &+ \cos\phi\cos\alpha \big)
    \label{eq:energy2}
\end{aligned}
\end{equation}

\noindent where $\phi$ and $\theta$ are the azimuthal and polar angles of the magnetization, $\alpha$ and $\beta$ are the azimuthal and polar angles of the applied field, and the Zeeman energy $-\mu_0 \mathbf{H} \cdot \mathbf{M}$ is expanded in spherical coordinates. Fig.\ \ref{fig:sup_free_energy_biaxial} shows the results of simulated magnetization loops in out-of-plane, and along the two in-plane symmetry axes. The main result is that, while in-plane anisotropy can give rise to hysteresis along the hard axis, to first order it cannot account for the observation that the coercive field in-plane is an order of magnitude larger than the coercive field out-of-plane. Note that for a general biaxial situation such behavior might be achieved in principle by having a very strong anisotropy in-plane relative to the energy cost of lying in-plane versus out-of-plane (ie, $K_3 >> K_1$). For hexagonal symmetry in particular, however, the $\cos6\theta$ dependence of the in-plane anisotropy drives the easy axis to switch from out-of-plane to in-plane since for angles $\theta = \frac{\pi(1 + 2n)}{6}$ for integer $n$ there is an energy \textit{saving} relative to other orientations. We emphasize that a full angle dependence on magnetization in-plane is needed to fully understand the hysteresis. One possibility is that higher order terms in the anisotropy can be used to accurately describe the data. Alternatively, such a study will reveal that more interesting physics is at play.

\section*{Structural characterization of doped samples}

\section*{Upper critical fields and estimation of coherence length in $Ni_{0.2}Ta_{4}Se_{8}$}

\begin{figure}[!htbp]
\centering
\includegraphics[scale=0.7]{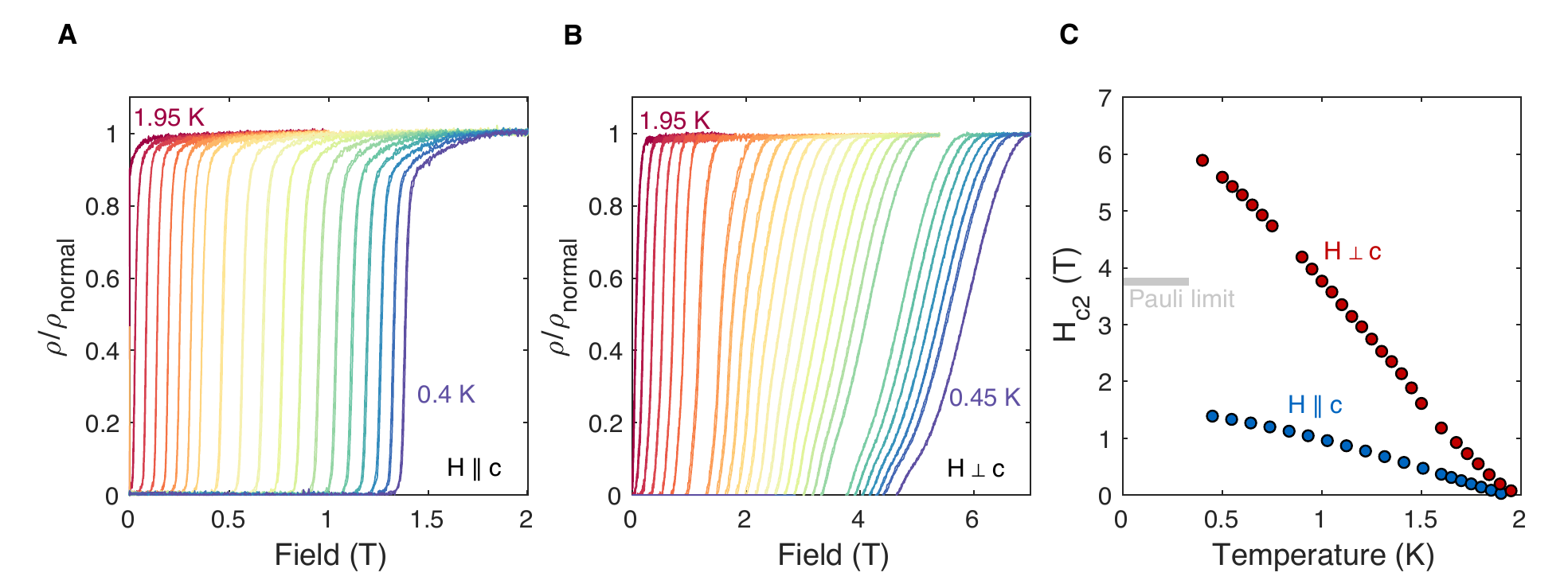}
\caption{{\bf Critical fields of Ni$_{0.2}$Ta$_{4}$Se$_{8}$} \textbf{A,B} Magnetoresistance for fields directed parallel and perpendicular to the crystallographic $c$-axis. $\textbf{C}$ Upper critical field, $H_{c2}$, determined via the point in the magnetoresistance curve with the steepest gradient. The Pauli limiting field is estimated for weak-coupling electrons by multiplying the superconducting transition temperature for this sample ($T_{c}$ = 1.95K) by 1.86.}
\label{fig:sup_criticalfields}
\end{figure}

The superconducting coherence length can be crudely estimated from the extrapolated value of the upper critical field at zero temperature, $H_{c2}(0)$, via the following formula
\begin{equation}
    \xi = \sqrt{\frac{\phi_{0}}{2\pi \mu_{0}H_{c2}(0)}},
\end{equation}
where $\phi_{0}$ is the flux quantum. The estimated zero-temperature critical fields (Fig.~\ref{fig:sup_criticalfields}C) are $H^{c}_{c2}(0) = 1.8 \pm 0.2$T and $H^{ab}_{c2}(0) = 7.0 \pm 0.2$T for the out-of-plane and in-plane field directions, respectively. This yields coherence lengths of $\xi^{c} = 13.5 \pm 1.5$ nm and  $\xi^{ab} = 6.9 \pm 0.2$ nm, implying that the material is a quasi two-dimensional superconductor. This is also evident in the strongly anisotropic diamagnetic shielding effect seen in Figs.~\ref{fig:superconductivity}B,C.

The superconductivity in this sample violates the Pauli limit when the external field is directed perpendicular to the crystallographic $c$-axis (Fig.~\ref{fig:sup_criticalfields}C). At present, it is not possible to say whether the Pauli limit violation observed in Fig.~\ref{fig:sup_criticalfields}C is indicative of the spin symmetry of the superconducting pairs, or results from the quasi two-dimensionality of the material in combination with spin-orbit coupling. 

We can also compare the coherence lengths listed here to the estimated mean free path of the carriers. The mean free path, $\l$, can be estimated via the Drude formula
\begin{equation}
    l = \frac{1}{\rho}\frac{1}{n e^{2}} \hbar k_{F},
\end{equation} where $\rho$ is the resistivity in the normal state at low temperature, $n$ is the carrier density, and $k_{F}$ is the Fermi momentum. Because the electron-like carriers appear to dominate both the Hall effect measurements, and are the main features in the Fermi surface, we here make a crude assumption that the main electron-like Fermi surface is the only set of electrons. The carrier density $n \approx 0.4 \times 10^{22}$cm$^{-3}$ can be estimated from Hall effect measurements, as shown in Fig.~\ref{fig:sup_Hall}B, where the dominant contribution to the Hall effect is the electron-like carriers. The DFT calculations shown in the main text suggest that $k_{F} \approx 0.25 \AA^{-1}$ for the main electron-like Fermi surface. The normal state resistivity (Fig.~\ref{fig:superconductivity}A) is about 125 $\mu\Omega$cm. Using these values, the mean free path for these carriers is approximately $l \approx 10$ nm. Note that here we report only one significant digit because of the error in estimating Fermi velocity and carrier density, and also the single carrier approximation described above, as well as the inherent assumptions built into the Drude formula. The value determined above is at least reliable within an order of magnitude. Thus, the intralayer mean free path is comparable to the superconducting coherence length inferred from critical field measurements ($\xi^{c}$), putting this superconductor in the dirty limit.

\newpage
\bibliographystyleSM{unsrt}
\bibliographySM{SMreferences}

\end{document}